\pdfoutput=1
\documentclass[prd,aps,twocolumn,preprintnumbers,amsmath,amssymb,floatfix,raggedbottom]{revtex4-2}
\usepackage{graphicx}
\usepackage[breaklinks,colorlinks=true,linkcolor=blue,citecolor=blue,urlcolor=blue]{hyperref}
\usepackage{bm}
\usepackage{physics}

\newcommand{\ri}{\mathrm{i}}
\newcommand{\re}{\mathrm{e}}
\newcommand{\Lag}{\mathcal{L}}
\newcommand{\Mcl}{M_{\mathrm{cl}}}
\newcommand{\ord}[1]{\mathcal{O}\!\left(#1\right)}
\newcommand{\redm}[3]{\langle #1 \Vert #2 \Vert #3\rangle}
\newcommand{\occ}{\mathrm{occ}}

\begin{document}

\title{Generalized parton distributions of the nucleon in the chiral quark soliton model: Sum over all quark states at nonzero skewness and momentum transfer}

\author{Josuke Minamiguchi}
\email{minamiguchi@nt.phys.s.u-tokyo.ac.jp}
\affiliation{Department of Physics, The University of Tokyo, 7-3-1 Hongo, Bunkyo-ku, Tokyo 113-0033, Japan}

\begin{abstract}
We compute the combinations $E_M=H^{u+d}+E^{u+d}$ and $H_E=H^{u-d}+tE^{u-d}/(4M_N^2)$ of unpolarized generalized parton distributions (GPDs) of the nucleon in the chiral quark soliton model at nonzero skewness $\xi$ and nonzero momentum transfer $t$.
The combinations contain the GPDs $H^{u-d}$ and $E^{u+d}$, which are subleading in the expansion in the inverse number of colors.
Both combinations are double sums over the quark states.
A multipole expansion of the light-cone operators makes the sums computable at every $\xi$ and every $t\le0$.
The contribution of the Dirac continuum is summed over the quark states, without an interpolation formula and without a gradient expansion.
The first moments satisfy the form-factor sum rules, and the second moments satisfy polynomiality.
\end{abstract}

\maketitle

\section{Introduction}\label{sec:intro}

Generalized parton distributions (GPDs) are matrix elements of quark operators on the light cone between nucleon states of different momenta~\cite{Mueller1994, Ji1997, Ji1997DVCS, Radyushkin1997}.
A GPD depends on the momentum fraction $x$, on the skewness $\xi$, and on the squared momentum transfer $t$.
GPDs contain information that neither the parton distribution functions nor the electromagnetic form factors contain.
For example, the second moment in $x$ of the GPD $E^q$ at $t=0$ enters the sum rule for the total angular momentum carried by the quarks of flavor $q$~\cite{Ji1997}.
Reviews are given in Refs.~\cite{Diehl2003, BelitskyRadyushkin2005}.

GPDs are accessible in hard exclusive processes such as deeply virtual Compton scattering (DVCS)~\cite{CollinsFrankfurtStrikman1997, Ji1997DVCS, Radyushkin1997}.
DVCS has been measured by several experiments~\cite{H1_2005, HERMES2012, Jo2015, Defurne2015, CLAS12_2023, COMPASS2019}.
At leading order in the strong coupling, only the values of a GPD at $x=\pm\xi$ and one integral of the GPD over $x$ enter the DVCS amplitude.
The data therefore do not determine the $x$ dependence of a GPD~\cite{Bertone2021}.
The analyses of the data use parametrizations of the $x$ dependence~\cite{VGG1999, GK2008, KM2010, GJSSY2023, KumerickiLiutiMoutarde2016}.
For the GPD $E^q$, even the forward limit is not known from deep-inelastic scattering~\cite{Diehl2003, Ossmann2005}.

Models of the nucleon give the dependence of GPDs on all three variables at a low normalization point~\cite{JiMelnitchoukSong1997, BoffiPasquiniTraini2003, ScopettaVento2003, MineoYangCheung2005, Petrov1998}.
One of these models is the chiral quark soliton model (CQSM)~\cite{Diakonov:1987ty, WakamatsuYoshiki, Christov:1995vm}.
The CQSM is a field theory of quarks in the limit of a large number of colors $N_c$.
In the CQSM the nucleon consists of $N_c$ quarks bound by a pion field, and the Dirac sea of the quarks is part of the nucleon.
The pion field is the saddle point of the effective action, which is proportional to $N_c$.
The nucleon is a quantum state of the collective rotation of the soliton.
The angular velocity $\Omega$ of this rotation is of order $1/N_c$.
The sum rules for the parton distributions of the model hold only if the contribution of the Dirac continuum is added to the contribution of the valence state~\cite{Diakonov1996, DPPPW1997, Pobylitsa1999, Wakamatsu2026}.
Each GPD has the two flavor combinations $u+d$ and $u-d$, for example $H^{u\pm d}=H^u\pm H^d$.
In general, at large $N_c$ one of the two combinations is larger than the other by one power of $N_c$~\cite{GPV2001, Penttinen2000, Ossmann2005}.
We call the larger combination leading in $1/N_c$ and the smaller combination subleading in $1/N_c$.
The leading combinations are $H^{u+d}$, $E^{u-d}$, $\tilde H^{u-d}$, and $\tilde E^{u-d}$.
The subleading combinations are $H^{u-d}$, $E^{u+d}$, $\tilde H^{u+d}$, and $\tilde E^{u+d}$.
In the CQSM this counting is the expansion in $\Omega$.
The expansion in $\Omega$ is not the $1/N_c$ expansion of QCD, but the leading order of each quantity is the same in the two.
The leading combinations are of zeroth order in $\Omega$.
Each of them is a single sum over the occupied quark states~\cite{Petrov1998, Penttinen2000, SchweitzerBoffiRadici2002, SchweitzerColliBoffi2003}.
The subleading combinations vanish at zeroth order in $\Omega$ by the symmetry of the static soliton.
At first order in $\Omega$ they are double sums over the quark states~\cite{Petrov1998, Ossmann2005}.
This first order is the leading term of each subleading combination, not a correction to a leading term.

Numerical results for the $x$ dependence at nonzero $\xi$ and $t$ exist for the leading combinations~\cite{Petrov1998, Penttinen2000, GPV2001}.
In these results, however, the contribution of the Dirac continuum is not summed over the quark states.
This contribution is obtained from an interpolation formula, the lowest order of an expansion of the quark propagator in the pion field~\cite{Petrov1998}.
Reference~\cite{Kim2025} estimated the subleading chiral-odd GPDs at nonzero $\xi$ and $t$.
The estimate keeps the ultraviolet-divergent terms of a gradient expansion of the contribution of the Dirac continuum and does not include the valence state.

The unpolarized subleading GPDs $H^{u-d}$ and $E^{u+d}$ naturally appear in the CQSM in two combinations~\cite{Ossmann2005}.
The combinations are $E_M=H^{u+d}+E^{u+d}$ and $H_E=H^{u-d}+tE^{u-d}/(4M_N^2)$, where $M_N$ is the nucleon mass.
The combinations $E_M$ and $H_E$ are of particular interest.
Their first moments are related to the nucleon isoscalar magnetic and isovector electric form factors, respectively.
The second moment of $E_M$ at $t=0$ gives twice the total angular momentum carried by quarks~\cite{Ji1997, Ossmann2005}.
Reference~\cite{Ossmann2005} derived the model expressions for $E_M$ and $H_E$ at nonzero $\xi$ and $t$.
The expressions are double sums over the quark states.
Each term of the double sums contains a matrix element of a light-cone operator between two quark states.
The same reference evaluated $E_M$ numerically only at $\xi=0$ and $t=0$.
To estimate observables, it modeled $E^u(x,\xi,t)$ and $E^d(x,\xi,t)$ with a double-distribution ansatz multiplied by form factors.
The ansatz takes the computed forward limit $E^{u+d}(x,0,0)$ as an input.
Reference~\cite{Ossmann2005} also states that the numerical technique used there does not allow the calculation of GPDs at all values of $\xi$ and $t$.
Moments in $x$ of $E_M$ and $H_E$ at nonzero $t$ have been computed from sums over the quark states~\cite{Christov:1995vm, WakamatsuNakakoji2006, Goeke2007}.
To our knowledge, no $x$ dependence at nonzero $\xi$ and $t$ has been computed with a sum over the states of the Dirac continuum.

In this paper we evaluate $E_M$ and $H_E$ directly at nonzero $\xi$ and $t$ from the expressions of Ref.~\cite{Ossmann2005}.
The calculation differs from the previous results at nonzero $\xi$ and $t$ in two respects.
First, the contribution of the Dirac continuum is summed over the quark states, without an interpolation formula and without a gradient expansion.
Second, the combinations are subleading in $1/N_c$, and the calculation includes both the valence state and the Dirac continuum.
At $t=0$ the symmetry of the soliton allows an average over the direction of the light cone.
The average brings the double sums to a spherically symmetric form, as in Eqs.~(50) and (51) of Ref.~\cite{Ossmann2005}.
A nonzero momentum transfer fixes a direction and forbids the average.
We expand the two functions of the one-particle operators in multipoles instead.
Selection rules in the grand spin restrict the multipoles, and the angular dependence of every matrix element follows in closed form.
The reduction holds for every $\xi$ and every $t\le0$.
The first moment of $E_M$ is proportional to the isoscalar magnetic form factor, and the first moment of $H_E$ is the isovector electric form factor.
The CQSM gives the two form factors as separate double sums.
Each first moment therefore provides a check at every $\xi$ and $t$.

Section~\ref{sec:model} gives the model, the collective quantization, and the regularization.
Section~\ref{sec:gpd} gives the definitions, the kinematics and the reduction of the double sums to the form evaluated.
The basis and the angular reduction are given in Appendix~\ref{app:basis} and Appendix~\ref{app:angular}.
Section~\ref{sec:numerics} gives the numerical procedure and the truncations.
Section~\ref{sec:results} presents the results, and Sec.~\ref{sec:discussion} discusses the results.

\section{The chiral quark soliton model}\label{sec:model}

\subsection{Effective action and the hedgehog soliton}

The chiral quark soliton model is defined by the Lagrangian~\cite{Diakonov:1987ty, Christov:1995vm}
\begin{equation}
  \Lag = \bar\psi\left(\ri\gamma^\mu\partial_\mu - MU^{\gamma_5}\right)\psi\,,
  \qquad U^{\gamma_5} = \exp\left(\ri\gamma_5\bm\tau\cdot\bm\pi/f_\pi\right)\,.
  \label{eq:lagrangian}
\end{equation}
Here $M$ is the constituent quark mass, $U \in SU(2)$ is the chiral field, and $N_c=3$ is the number of colors.
Integrating out the quarks gives the effective action
\begin{equation}
  \Gamma[U] = -\ri N_c\Tr\ln\left(\ri\gamma^\mu\partial_\mu - MU^{\gamma_5}\right)\,.
  \label{eq:effective-action}
\end{equation}
At large $N_c$ the functional integral over $U$ is dominated by the saddle point of $\Gamma[U]$.
This saddle point is the hedgehog
\begin{equation}
    U_c^{\gamma_5}(\bm r) = \exp\left[\ri\gamma_5\bm\tau\cdot\bm{e}_rF(r)\right]\,.
  \label{eq:hedgehog}
\end{equation}
The profile satisfies $F(0)=-\pi$ and $F(\infty)=0$, and $\bm{e}_r$ is the radial unit vector.
The one-particle Hamiltonian is
\begin{equation}
  H_c = -\ri\gamma^0\gamma^k\partial_k + \gamma^0MU_c^{\gamma_5}\,,
  \qquad H_c\ket{n} = E_n\ket{n}\,.
  \label{eq:dirac-spectrum}
\end{equation}
The hedgehog is invariant under simultaneous rotations in space and in isospin.
The Hamiltonian $H_c$ therefore commutes with the grand-spin operator $\hat{\bm K} = \hat{\bm J} + \hat{\bm T}$, where $\hat{\bm J} = \hat{\bm L} + \hat{\bm S}$.
The operators $\hat{\bm L}$, $\hat{\bm S}$, and $\hat{\bm T}$ are the orbital angular momentum, the spin, and the isospin of the quark.
The Hamiltonian $H_c$ also commutes with the parity operator $\hat\Pi$.
The eigenstates of $H_c$ can therefore be chosen as eigenstates of $\hat{\bm K}^2$, $\hat K^3$, and $\hat\Pi$.
We write the eigenvalues as $K(K+1)$, $K^3$, and $\Pi=\pm1$, and we call $K$ the grand spin.

The sector $(K,\Pi) = (0,+)$ contains one discrete state with $-M<E<M$, the valence state, of energy $E_{\mathrm{val}}$.
The negative Dirac continuum consists of the states with $E<0$ other than the valence state.
The occupied states are the states with $E\le E_{\mathrm{val}}$, that is, the valence state and the negative Dirac continuum.
All other states are the non-occupied states.
Below, $\occ$ labels the occupied states.
The profile $F(r)$ minimizes the classical soliton mass
\begin{equation}
  \Mcl = N_c\sum_{n\in\occ}E_n - N_c\sum_{n\in\occ}E_n\Big|_{F\equiv0}\,.
  \label{eq:Mcl}
\end{equation}
The mass $\Mcl$ is regularized as in Sec.~\ref{sec:reg}.
We obtain $F(r)$ by self-consistent iteration~\cite{WakamatsuKubota, PVdouble}.

\subsection{Collective quantization}

Spin and isospin are generated by a slow collective rotation $A(t)\in SU(2)$,
\begin{equation}
  U^{\gamma_5}(\bm r,t) = A(t)U_c^{\gamma_5}(\bm r)A^\dagger(t)\,,
  \qquad A^\dagger\dot A = \frac{\ri}{2}\Omega^a\tau^a\,.
  \label{eq:collective-rotation}
\end{equation}
The moment of inertia is~\cite{Pobylitsa1999}
\begin{equation}
  I = \frac{N_c}{6}\sum_{n\in\occ}\ \sum_{m:\,E_m\neq E_n}
      \frac{\mel{n}{\tau^a}{m}\mel{m}{\tau^a}{n}}{E_m - E_n}\,.
  \label{eq:inertia}
\end{equation}
The form of Eq.~\eqref{eq:inertia} and the form with $m$ restricted to the non-occupied states have the same value~\cite{WakamatsuYoshiki, Ossmann2005}.
The subtraction of Sec.~\ref{sec:reg} acts on the two forms differently.
We follow Ref.~\cite{Pobylitsa1999} and evaluate $I$ from Eq.~\eqref{eq:inertia}.
The double sums of Sec.~\ref{sec:spectral} have the same range of $m$.
For $J=\tfrac12$, quantization of Eq.~\eqref{eq:collective-rotation} gives
\begin{equation}
  M_N = \Mcl + \frac{J(J+1)}{2I} = \Mcl + \frac{3}{8I}\,.
  \label{eq:MN}
\end{equation}
Equation~(12) of Ref.~\cite{Ossmann2005} defines the symbol $M_N$ used there as the static soliton energy.
This energy is $\Mcl$ of Eq.~\eqref{eq:Mcl}.
All prefactors and all kinematics below therefore use $\Mcl$.
The difference between $\Mcl$ and $M_N$ is of relative order $N_c^{-2}$.

\subsection{Regularization}\label{sec:reg}

The quark loop of Eq.~\eqref{eq:effective-action} is ultraviolet divergent.
We regularize the effective action by the double Pauli--Villars subtraction~\cite{PVdouble}
\begin{equation}
  \Gamma^{\mathrm{reg}}[U] = \Gamma_M[U]
    - \sum_{i=1}^{2}c_i\,\Gamma_{\Lambda_i}[U]\,.
  \label{eq:pvaction}
\end{equation}
The subscript of $\Gamma$ is the quark mass in $H_c$ of Eq.~\eqref{eq:dirac-spectrum}.
The divergences cancel if
\begin{equation}
  \sum_ic_i\Lambda_i^2 = M^2\,,\qquad \sum_ic_i\Lambda_i^4 = M^4\,.
  \label{eq:pvconditions}
\end{equation}
These two conditions determine $c_1$ and $c_2$ from $\Lambda_1$ and $\Lambda_2$.
The two masses are fixed by the pion decay constant and the chiral condensate~\cite{PVdouble},
\begin{align}
  f_\pi^2 &= \frac{N_cM^2}{4\pi^2}\sum_ic_i
             \left(\frac{\Lambda_i}{M}\right)^{2}\ln\frac{\Lambda_i^2}{M^2}\,,
  \label{eq:fpi-condition} \\
  \expval{\bar qq} &= \frac{N_cM^3}{2\pi^2}\sum_ic_i
             \left(\frac{\Lambda_i}{M}\right)^{4}\ln\frac{\Lambda_i^2}{M^2}\,.
  \label{eq:condensate-condition}
\end{align}
The constituent mass $M$ is then the only free parameter.
Because Eq.~\eqref{eq:pvaction} regularizes the action, the quantities derived from the action inherit the subtraction of Eq.~\eqref{eq:pvaction}.
Following Ref.~\cite{Ossmann2005}, however, we apply the subtraction to the sea contribution alone.
Let $\mathcal Q^{\mathrm{sea}}_\mu[F]$ denote the sea contribution to a quantity.
This sea contribution is computed with the quark mass $\mu$ and the profile $F$.
We evaluate
\begin{equation}
  \begin{aligned}
  \mathcal Q^{\mathrm{sea}}_{\mathrm{reg}}
   &= \left\{\mathcal Q^{\mathrm{sea}}_M[F]
      - \mathcal Q^{\mathrm{sea}}_M[0]\right\} \\
   &\quad - \sum_{i=1}^{2}c_i
     \left\{\mathcal Q^{\mathrm{sea}}_{\Lambda_i}[F]
      - \mathcal Q^{\mathrm{sea}}_{\Lambda_i}[0]\right\}\,.
  \end{aligned}
  \label{eq:Qreg}
\end{equation}
The valence contribution, computed with the quark mass $M$, is added to $\mathcal Q^{\mathrm{sea}}_{\mathrm{reg}}$.
We use Eq.~\eqref{eq:pvaction} at every stage of the calculation, including the self-consistent profile.

\section{Generalized parton distributions in the model}\label{sec:gpd}

\subsection{Definitions}

The unpolarized generalized parton distributions (GPDs) are defined by~\cite{Ji1997, Ossmann2005}
\begin{widetext}
\begin{align}
  \int\frac{dz}{2\pi}\,\re^{\ri zx}
    &\mel{P',S_3'}{\bar\psi_q\!\left(-\tfrac{zn}{2}\right)
      n_\mu\gamma^\mu\left[-\tfrac{zn}{2},\tfrac{zn}{2}\right]
      \psi_q\!\left(\tfrac{zn}{2}\right)}{P,S_3}
  \nonumber \\
  &= H^q(x,\xi,t)\,\bar u(P',S_3')\,n_\mu\gamma^\mu\,u(P,S_3)
   + E^q(x,\xi,t)\,\bar u(P',S_3')\,
     \frac{\ri\sigma^{\mu\nu}n_\mu\Delta_\nu}{2M_N}\,u(P,S_3)\,.
  \label{eq:defGPD}
\end{align}
\end{widetext}
In Eq.~\eqref{eq:defGPD}, $u(P,S_3)$ is the nucleon spinor and $[z_1,z_2]$ is the gauge link.
The light-cone vector $n$ satisfies $n\cdot(P'+P)=2$.
The momentum transfer is $\Delta^\mu = (P'-P)^\mu$, and the squared momentum transfer is $t=\Delta^2$.
The relation $n\cdot\Delta = -2\xi$ defines the skewness $\xi$.
The antiquark distributions are $H^{\bar q}(x,\xi,t) = -H^q(-x,\xi,t)$ and $E^{\bar q}(x,\xi,t) = -E^q(-x,\xi,t)$.
The first moments are $\int dx\,H^q = F_1^q(t)$ and $\int dx\,E^q = F_2^q(t)$.
We write
\[
H^{u\pm d}=H^u\pm H^d,\qquad
E^{u\pm d}=E^u\pm E^d.
\]
We compute the two combinations~\cite{Ossmann2005}
\begin{align}
E_M(x,\xi,t)
&\equiv H^{u+d}(x,\xi,t)+E^{u+d}(x,\xi,t),
\\
H_E(x,\xi,t)
&\equiv H^{u-d}(x,\xi,t)
+\frac{t}{4M_N^2}E^{u-d}(x,\xi,t).
\end{align}
The first combination is of order $N_c^2$, whereas the second is of order $N_c$.
The first moment of $E_M$ is the isoscalar magnetic form factor, and the first moment of $H_E$ is the isovector electric form factor\footnote{Equation~(23) of Ref.~\cite{Ossmann2005}, as printed, interchanges $G_E$ and $G_M$.
We use the standard definitions $G_E^q = F_1^q + (t/4M_N^2)F_2^q$ and $G_M^q = F_1^q + F_2^q$.
The sum rules in Eq.~(30) of Ref.~\cite{Ossmann2005}, which are Eqs.~\eqref{eq:sumEM} and \eqref{eq:sumHE} here, hold with these definitions.},
\begin{align}
  \int dx\,E_M(x,\xi,t) &= 3\,G_M^{p+n}(t)\,,
  \label{eq:sumEM} \\
  \int dx\,H_E(x,\xi,t) &= G_E^{p-n}(t)\,.
  \label{eq:sumHE}
\end{align}
We write $G_M^{p+n} = G_M^p + G_M^n$ and $G_E^{p-n} = G_E^p - G_E^n$ for the form factors of the proton and the neutron.
For form factors the isoscalar combination is the sum over the proton and the neutron.
For GPDs the isoscalar combination is the sum over the quark flavors.
The two conventions give the factor three in Eq.~\eqref{eq:sumEM}.
In the model the right-hand sides of Eqs.~\eqref{eq:sumEM} and \eqref{eq:sumHE} are the double sums~\cite{Ossmann2005}
\begin{align}
    3G_M^{p+n}(t) &= \frac{\Mcl N_c}{2I\bm\Delta^2}
     \sum_{n\in\occ}\ \sum_{m:\,E_m\neq E_n}
     \frac{\ri\,\epsilon^{abc}\Delta^a}{E_n-E_m}
  \nonumber \\
   &\qquad\times
     \mel{n}{\tau^b}{m}
     \mel{m}{\gamma^0\gamma^c\re^{\ri\bm\Delta\cdot\hat{\bm X}}}{n}\,,
  \label{eq:GMdef} \\
  G_E^{p-n}(t) &= -\frac{N_c}{6I}\sum_{n\in\occ}\ \sum_{m:\,E_m\neq E_n}
     \frac{1}{E_n-E_m}
  \nonumber \\
   &\qquad\times
     \mel{n}{\tau^a}{m}\mel{m}{\tau^a\re^{\ri\bm\Delta\cdot\hat{\bm X}}}{n}\,,
  \label{eq:GEdef}
\end{align}
where $\hat{\bm X}$ is the position operator.
The range of $m$ is that of Eq.~\eqref{eq:inertia}, and both sums are regularized as $I$.
At $t=0$ the factor $\re^{\ri\bm\Delta\cdot\hat{\bm X}}$ equals $1$, and the right-hand side of Eq.~\eqref{eq:GEdef} becomes the right-hand side of Eq.~\eqref{eq:inertia} divided by $I$.
The form factor $G_E^{p-n}(0)$ therefore equals $1$ exactly, also after the regularization.
This value provides a check of the computed $H_E$.

\subsection{Kinematics}\label{sec:kinematics}

In the large-$N_c$ kinematics, the nucleon mass is $\ord{N_c}$, the spatial momentum transfer $\Delta^i$ is $\ord{N_c^0}$, and the energy transfer $\Delta^0$ is $\ord{N_c^{-1}}$~\cite{Ossmann2005}.
The average momentum $\bar P = (P+P')/2$ therefore has the components $\bar P^0 = M_N+\ord{N_c^{-1}}$ and $\bar P^3 = \ord{N_c^0}$.
With $n^\mu = (1,0,0,-1)/\Mcl$, the definitions of $\xi$ and $t$ below Eq.~\eqref{eq:defGPD} give
\begin{equation}
  \xi = -\frac{\Delta^3}{2\Mcl} + \ord{N_c^{-2}}\,,\qquad
  t = -\bm\Delta^2 + \ord{N_c^{-2}}\,.
  \label{eq:kinematics}
\end{equation}
Because $\Delta^3 = \ord{N_c^0}$, Eq.~\eqref{eq:kinematics} gives $\xi = \ord{N_c^{-1}}$.
We drop the $\ord{N_c^{-2}}$ terms and use
\begin{equation}
  \Delta^3 = -2\Mcl\,\xi\,,\qquad
  \Delta_\perp^2 = -t - 4\Mcl^2\xi^2\,.
  \label{eq:delta3}
\end{equation}
The model kinematics is restricted to $-t\ge4\Mcl^2\xi^2$ by $\Delta_\perp^2\ge0$.

The hedgehog is not invariant under spatial rotations alone, so the matrix elements are taken in a fixed frame.
The light-cone direction is $\bm{e}_z$ and
\begin{equation}
  \bm\Delta = (\Delta_\perp,\,0,\,\Delta^3)\,,\qquad
  \Delta^3 = |\bm\Delta|\cos\Theta_\Delta\,.
  \label{eq:frame}
\end{equation}
For given $\xi$ and $t$, Eqs.~\eqref{eq:delta3} and \eqref{eq:frame} give
\begin{equation}
  |\bm\Delta| = \sqrt{-t}\,,\qquad
  \cos\Theta_\Delta = \frac{-2\Mcl\,\xi}{\sqrt{-t}}\,.
  \label{eq:kin-input}
\end{equation}
The forward limit is $\bm\Delta\to0$, that is $\xi\to0$ and $t\to0$.

\subsection{Spectral representation}\label{sec:spectral}

The two distributions are double sums over the single-quark states of Eq.~\eqref{eq:dirac-spectrum}~\cite{Ossmann2005},
\begin{widetext}
\begin{align}
  H_E(x,\xi,t) &= -\frac{\Mcl N_c}{12I}\int\frac{dz^0}{2\pi}\,\re^{\ri z^0x\Mcl}
    \biggl[\biggl\{\sum_{\substack{n\in\occ,\,m=\mathrm{all}\\ E_m\neq E_n}}\re^{-\ri z^0E_n}
      -\sum_{\substack{n=\mathrm{all},\,m\in\occ\\ E_m\neq E_n}}\re^{-\ri z^0E_m}\biggr\}
      \frac{1}{E_n-E_m}
  \nonumber \\
  &\qquad+\frac{\partial}{\partial(x\Mcl)}
       \sum_{\substack{n\in\occ,\,m=\mathrm{all}\\ E_m\neq E_n}}\re^{-\ri z^0E_n}\biggr]
  \nonumber \\
  &\qquad\times\mel{n}{\tau^a}{m}
   \mel{m}{\tau^a\left(1+\gamma^0\gamma^3\right)
     \re^{-\ri z^0\hat p^3/2}\re^{\ri\bm\Delta\cdot\hat{\bm X}}
     \re^{-\ri z^0\hat p^3/2}}{n}\,,
  \label{eq:oss28} \\[1ex]
  E_M(x,\xi,t) &= \frac{\ri\Mcl^2N_c}{2I}\int\frac{dz^0}{2\pi}\,\re^{\ri z^0x\Mcl}
    \biggl[\biggl\{\sum_{\substack{n\in\occ,\,m=\mathrm{all}\\ E_m\neq E_n}}\re^{-\ri z^0E_n}
      -\sum_{\substack{n=\mathrm{all},\,m\in\occ\\ E_m\neq E_n}}\re^{-\ri z^0E_m}\biggr\}
      \frac{1}{E_n-E_m}
  \nonumber \\
  &\qquad+\frac{\partial}{\partial(x\Mcl)}
       \sum_{\substack{n\in\occ,\,m=\mathrm{all}\\ E_m\neq E_n}}\re^{-\ri z^0E_n}\biggr]
  \nonumber \\
  &\qquad\times\mel{n}{\tau^b}{m}
   \mel{m}{\left(1+\gamma^0\gamma^3\right)
     \re^{-\ri z^0\hat p^3/2}\frac{\epsilon^{3ab}\Delta^a}{\Delta_\perp^2}
     \re^{\ri\bm\Delta\cdot\hat{\bm X}}\re^{-\ri z^0\hat p^3/2}}{n}\,.
  \label{eq:oss29}
\end{align}
\end{widetext}
The operator $\hat p^3$ is the longitudinal momentum operator.
Equations~\eqref{eq:oss28} and \eqref{eq:oss29} are the leading terms of the $1/N_c$ expansions of $H_E$ and $E_M$.
The two equations differ in three places only: the prefactor, the isospin structure, and the factor $\epsilon^{3ab}\Delta^a/\Delta_\perp^2$.

\subsection{Reduction to the form evaluated}\label{sec:reduction}

Steps (i)--(iv) are identities, step (v) is the only approximation of this section, and step (vi) assembles the form evaluated.
Section~\ref{sec:numerics} describes the truncations of the basis and of the multipole sums.

\emph{(i) The $z^0$ integral and the two operator orderings.}
This step carries out the $z^0$ integral of Eqs.~\eqref{eq:oss28} and \eqref{eq:oss29} and brings the two sums of each equation to the same range of $n$ and $m$.
With $\Delta^3=-2\Mcl\xi$ of Eq.~\eqref{eq:delta3},
\begin{equation}
  \re^{-\ri z^0\hat p^3/2}\re^{\ri\bm\Delta\cdot\hat{\bm X}}
  \re^{-\ri z^0\hat p^3/2}
  = \re^{\ri\bm\Delta\cdot\hat{\bm X}}\re^{-\ri z^0\hat p^3}
    \re^{\ri z^0\Mcl\xi}\,.
  \label{eq:bch}
\end{equation}
The $z^0$ integral of the exponential factors in Eqs.~\eqref{eq:oss28} and \eqref{eq:oss29} then gives a delta function.
The one-particle operator is
\begin{equation}
  \hat{\mathcal O}(E) = \left(1+\gamma^0\gamma^3\right)
    \re^{\ri\bm\Delta\cdot\hat{\bm X}}\,
    \delta\!\left((x+\xi)\Mcl - E - \hat p^3\right)\,.
  \label{eq:Odef}
\end{equation}

In Eq.~\eqref{eq:oss29}, the factor $\epsilon^{3ab}\Delta^a/\Delta_\perp^2$ is contracted with $\tau^b$ in the first matrix element, so we move the factor to the first matrix element.
In the frame of Eq.~\eqref{eq:frame}, $\bm\Delta$ has no $y$ component, so
\begin{equation}
  \frac{\epsilon^{3ab}\Delta^a\tau^b}{\Delta_\perp^2}
   = \frac{\tau^y}{\Delta_\perp}\,,\qquad
  \tau^y = \frac{\ri}{\sqrt2}\left(\tau_{+1}+\tau_{-1}\right)\,,
  \label{eq:tauy}
\end{equation}
where $\tau_{\pm1} = \mp(\tau^x\pm\ri\tau^y)/\sqrt2$ and $\tau_0 = \tau^z$ are the spherical components of $\bm\tau$.
For $H_E$, we write the contraction of the two isospin matrices as $\tau^a\tau^a=\sum_q(-1)^q\tau_q\tau_{-q}$ with $q=0,\pm1$.
In the second sum of Eqs.~\eqref{eq:oss28} and \eqref{eq:oss29} the occupied state is the intermediate state $m$.
We rename the summation labels in the second sum and use $1/(E_m-E_n) = -1/(E_n-E_m)$.
Both sums then run over $n\in\occ$ and $m=\mathrm{all}$.
With Eq.~\eqref{eq:tauy}, the contraction over $b$ in Eq.~\eqref{eq:oss29} becomes a sum over $q=\pm1$, and Eqs.~\eqref{eq:oss28} and \eqref{eq:oss29} become
\begin{align}
  H_E &= -\frac{\Mcl N_c}{12I}
   \sum_{\substack{n\in\occ,\,m=\mathrm{all}\\ E_m\neq E_n}}
  \nonumber \\
   &\qquad\times
   \left[\frac{\mathcal T^{H_E}_{\mathrm I}+\mathcal T^{H_E}_{\mathrm{II}}}{E_n-E_m}
     + \frac{1}{\Mcl}\frac{\partial}{\partial x}\mathcal T^{H_E}_{\mathrm I}\right]\,,
  \label{eq:two-orders-HE} \\
    E_M &= -\frac{\Mcl^2N_c}{2I}\frac{1}{\sqrt2\,\Delta_\perp}\sum_{q=\pm1}
   \sum_{\substack{n\in\occ,\,m=\mathrm{all}\\ E_m\neq E_n}}
  \nonumber \\
   &\qquad\times
   \left[\frac{\mathcal T^{E_M}_{\mathrm I,q}+\mathcal T^{E_M}_{\mathrm{II},q}}{E_n-E_m}
     + \frac{1}{\Mcl}\frac{\partial}{\partial x}\mathcal T^{E_M}_{\mathrm I,q}\right]\,.
  \label{eq:two-orders-EM}
\end{align}
In both equations, the term with the subscript I comes from the first sum, and the term with the subscript II comes from the second sum.
For $H_E$,
\begin{align}
  \mathcal T^{H_E}_{\mathrm I} &=
    \sum_q(-1)^q\mel{n}{\tau_q}{m}\mel{m}{\tau_{-q}\hat{\mathcal O}(E_n)}{n}\,,
  \nonumber \\
  \mathcal T^{H_E}_{\mathrm{II}} &=
    \sum_q(-1)^q\mel{m}{\tau_q}{n}\mel{n}{\tau_{-q}\hat{\mathcal O}(E_n)}{m}\,.
  \label{eq:T-HE}
\end{align}
For $E_M$,
\begin{align}
  \mathcal T^{E_M}_{\mathrm I,q} &=
    \mel{n}{\tau_q}{m}\mel{m}{\hat{\mathcal O}(E_n)}{n}\,,
  \nonumber \\
  \mathcal T^{E_M}_{\mathrm{II},q} &=
    \mel{m}{\tau_q}{n}\mel{n}{\hat{\mathcal O}(E_n)}{m}\,.
  \label{eq:T-EM}
\end{align}

\emph{(ii) Multipole expansions.}
This step expands the operator $\hat{\mathcal O}(E_n)$ in the terms of Eqs.~\eqref{eq:T-HE} and \eqref{eq:T-EM}.
This expansion is what makes the double sums computable at every $\xi$ and every $t\le0$.
The two operators in Eq.~\eqref{eq:Odef} are expanded in the Racah-normalized spherical harmonics $C^\kappa_\mu = \sqrt{4\pi/(2\kappa+1)}\,Y_{\kappa\mu}$.
Here $Y_{\kappa\mu}$ are the spherical harmonics in the Condon--Shortley convention, and $\bm{e}_p$ and $\bm{e}_\Delta$ are the directions of $\bm p$ and $\bm\Delta$.
The expansions are
\begin{align}
  &\re^{\ri\bm\Delta\cdot\bm r} = \sum_{L\mu}\ri^L(2L+1)
     \big[C^L_\mu(\bm{e}_\Delta)\big]^{*}j_L(|\bm\Delta|r)\,C^L_\mu(\bm{e}_r)\,,
  \label{eq:rayleigh} \\
  &\delta(A-p\,\bm{e}_p\!\cdot\!\bm{e}_z) = \sum_\lambda\frac{2\lambda+1}{2p}
     P_\lambda(A/p)\Theta(p-|A|)\,C^\lambda_0(\bm{e}_p)\,.
  \label{eq:lcmultipole}
\end{align}
In Eq.~\eqref{eq:lcmultipole}, $A$ denotes $(x+\xi)\Mcl - E_n$ and $p$ denotes $|\bm p|$.
The functions $j_L$ are the spherical Bessel functions, $P_\lambda$ are the Legendre polynomials, and $\Theta$ is the step function.
The azimuth of $\bm\Delta$ vanishes in the frame of Eq.~\eqref{eq:frame}, so $C^L_\mu(\bm{e}_\Delta)$ is real.
Each term of Eqs.~\eqref{eq:T-HE} and \eqref{eq:T-EM} starts and ends with the same state, so the spherical components of its operators sum to zero.
Only $\mu=0$ therefore contributes to $H_E$, and only $\mu=-q$ contributes to $E_M$.
Because $|\mu|\le L$, the sum over $L$ for $E_M$ starts at $L=1$.
For $E_M$, the factor $1/\Delta_\perp$ of Eq.~\eqref{eq:two-orders-EM} is absorbed with
\begin{equation}
\frac{C^L_{-q}(\bm{e}_\Delta)}{\Delta_\perp} =
q\,\frac{P_L'(\cos\Theta_\Delta)}{\sqrt{L(L+1)}\,|\bm\Delta|}\,.
  \label{eq:absorb}
\end{equation}
The coefficients $\mathcal E^{H_E}_L$ and $\mathcal E^{E_M}_{Lq}$ collect the factors of Eqs.~\eqref{eq:rayleigh} and \eqref{eq:absorb} that do not depend on $\bm r$.
The weights $w^{H_E}_L$ and $w^{E_M}_L$ collect the radial factors,
\begin{equation}
  \begin{aligned}
  (H_E,\ L\ge0)\quad \mathcal E^{H_E}_L &= \ri^L(2L+1)P_L(\cos\Theta_\Delta)\,, \\
  w^{H_E}_L(r) &= j_L(|\bm\Delta|r)\,, \\
  (E_M,\ L\ge1)\quad \mathcal E^{E_M}_{Lq} &= q\,\ri^L(2L+1)\,
     \frac{P_L'(\cos\Theta_\Delta)}{\sqrt{L(L+1)}}\,, \\
  w^{E_M}_L(r) &= \frac{j_L(|\bm\Delta|r)}{|\bm\Delta|}\,.
  \end{aligned}
  \label{eq:multipole-coeff}
\end{equation}
We insert Eqs.~\eqref{eq:Odef}, \eqref{eq:rayleigh}, and \eqref{eq:lcmultipole} into Eqs.~\eqref{eq:T-HE} and \eqref{eq:T-EM} and keep the contributing value of $\mu$.
With the coefficients of Eq.~\eqref{eq:multipole-coeff}, the insertion gives
\begin{widetext}
\begin{align}
  \mathcal T^{H_E}_{\mathrm I}
  &= \sum_q(-1)^q\sum_{L\ge0}\mathcal E^{H_E}_L\sum_\lambda
     \mel{n}{\tau_q}{m}
  \nonumber \\
  &\quad\times
     \mel{m}{\tau_{-q}\left(1+\gamma^0\gamma^3\right)w^{H_E}_L(r)\,C^L_0(\bm{e}_r)\,
       \frac{2\lambda+1}{2p}\,P_\lambda(A/p)\,\Theta(p-|A|)\,
       C^\lambda_0(\bm{e}_p)}{n}\,,
  \label{eq:TI-HE} \\
  \frac{\mathcal T^{E_M}_{\mathrm I,q}}{\Delta_\perp}
  &= \sum_{L\ge1}\mathcal E^{E_M}_{Lq}\sum_\lambda
     \mel{n}{\tau_q}{m}
  \nonumber \\
  &\quad\times
     \mel{m}{\left(1+\gamma^0\gamma^3\right)w^{E_M}_L(r)\,C^L_{-q}(\bm{e}_r)\,
       \frac{2\lambda+1}{2p}\,P_\lambda(A/p)\,\Theta(p-|A|)\,
       C^\lambda_0(\bm{e}_p)}{n}\,.
  \label{eq:TI-EM}
\end{align}
\end{widetext}
Inside the matrix elements, $r$ and $\bm{e}_r$ are the magnitude and the direction of $\hat{\bm X}$.
The symbols $p$ and $\bm{e}_p$ are the magnitude and the direction of the momentum operator.
The factor $1/\Delta_\perp$ is that of Eq.~\eqref{eq:two-orders-EM}.
Both weights are finite at $\Delta_\perp=0$ and at $|\bm\Delta|=0$.
In the forward limit the weight of $H_E$ tends to $\delta_{L0}$, so only $L=0$ contributes to $H_E$.
The weight of $E_M$ tends to $r/3$ for $L=1$ and to zero for $L\ge2$, so only $L=1$ contributes to $E_M$.

\emph{(iii) Basis expansion.}
This step inserts the Kahana--Ripka basis~\cite{KahanaRipka} into the second matrix element of Eqs.~\eqref{eq:TI-HE} and \eqref{eq:TI-EM}.
The eigenstates of a sector $(K,\Pi)$ come in multiplets of $2K+1$ states with the same energy.
We write $K_n$ and $K_n^3$ for the grand spin of the state $n$ and its projection, and $K_m$ and $K_m^3$ for those of the state $m$.
The eigenstates are expanded as
\begin{equation}
  \ket{n} = \sum_\beta V_{\beta n}\ket{\beta K_n^3}\,,
  \label{eq:expansion}
\end{equation}
with real $V$.
The basis states are the solutions of Eq.~\eqref{eq:dirac-spectrum} at $F\equiv0$ in a spherical box of radius $D$.
The label $\beta$ stands for the sector $(K_\beta,\Pi_\beta)$, the total angular momentum $j=K_\beta\pm\tfrac12$, the sign of the energy, and the discrete momentum $p_\beta$ (Appendix~\ref{app:basis}).
The coefficient $V_{\beta n}$ does not depend on $K_n^3$ and vanishes unless the basis state and the eigenstate belong to the same sector.
In position space
\begin{equation}
  \begin{aligned}
  \braket{\bm r}{\beta K^3} &=
    \begin{pmatrix}
      \mathfrak c^{(\mathrm{up})}_\beta\,\chi^{(\mathrm{up})}_\beta(\bm r) \\
      \mathfrak c^{(\mathrm{dn})}_\beta\,\chi^{(\mathrm{dn})}_\beta(\bm r)
    \end{pmatrix}\,, \\
  \chi^{(c)}_\beta(\bm r) &= j_{l_c}(p_\beta r)\,
    \mathcal Y_{K_\beta K^3 l_c j}(\bm{e}_r)\,, \\
  \mathcal Y_{KK^3lj}
   &= \Big[\big[Y_l\otimes\zeta\big]_{j}\otimes\eta\Big]_{KK^3}\,,
  \end{aligned}
  \label{eq:coupled}
\end{equation}
where $\zeta$ and $\eta$ are the spin and isospin spinors of the quark~\cite{WakamatsuYoshiki}, $l_c$ is the orbital angular momentum of the Dirac component $c=\mathrm{up},\mathrm{dn}$, and $\mathfrak c^{(c)}_\beta$ are the coefficients of the free Dirac spinor in the box (Appendix~\ref{app:basis}).
We write $\ket{KK^3\,lj}$ for the angular state with $\braket{\bm{e}_r}{KK^3\,lj} = \mathcal Y_{KK^3lj}(\bm{e}_r)$.
The factor $1+\gamma^0\gamma^3$ of Eqs.~\eqref{eq:TI-HE} and \eqref{eq:TI-EM} is split into blocks.
In the Dirac representation
\begin{equation}
  1+\gamma^0\gamma^3 = \begin{pmatrix} 1 & \sigma_0 \\ \sigma_0 & 1\end{pmatrix}\,.
  \label{eq:dirac-split}
\end{equation}
The rows and the columns refer to the upper (up) and the lower (dn) Dirac components.
Here $\sigma_0=\sigma_z$ is the $\mu=0$ spherical component of the Pauli matrices $\bm\sigma$.
We write $\mathcal D=0$ for the two diagonal blocks and $\mathcal D=1$ for the two off-diagonal blocks.
The set $\mathcal P_{\mathcal D}$ contains the pairs of components of the blocks: $\mathcal P_{\mathcal D=0} = \{(\mathrm{up},\mathrm{up}),(\mathrm{dn},\mathrm{dn})\}$ and $\mathcal P_{\mathcal D=1} = \{(\mathrm{up},\mathrm{dn}),(\mathrm{dn},\mathrm{up})\}$.
We insert Eqs.~\eqref{eq:expansion}, \eqref{eq:coupled}, and \eqref{eq:dirac-split} into the second matrix element of Eqs.~\eqref{eq:TI-HE} and \eqref{eq:TI-EM}.
The insertion gives
\begin{widetext}
\begin{align}
  &\mel{m}{\tau_{-q}\left(1+\gamma^0\gamma^3\right)w^{H_E}_L(r)\,
       C^L_0(\bm{e}_r)\,\frac{2\lambda+1}{2p}\,P_\lambda(A/p)\,\Theta(p-|A|)\,
       C^\lambda_0(\bm{e}_p)}{n}
  \nonumber \\
  &\qquad= \sum_{\beta'\beta}V_{\beta'm}V_{\beta n}
     \sum_{\mathcal D}\sum_{(c',c)\in\mathcal P_{\mathcal D}}
     \big(\mathfrak c^{(c')}_{\beta'}\big)^{*}\mathfrak c^{(c)}_{\beta}
     \frac{2\lambda+1}{2p_\beta}\,P_\lambda(A/p_\beta)\,
     \Theta(p_\beta-|A|)
  \nonumber \\
  &\qquad\qquad\times
     \mel{\chi^{(c')}_{\beta'}}{\tau_{-q}\,(\sigma_0)^{\mathcal D}\,
       w^{H_E}_L(r)\,C^L_0(\bm{e}_r)\,C^\lambda_0(\bm{e}_p)}
         {\chi^{(c)}_{\beta}}\,,
  \label{eq:afterbasis} \\
  &\mel{m}{\left(1+\gamma^0\gamma^3\right)w^{E_M}_L(r)\,
       C^L_{-q}(\bm{e}_r)\,\frac{2\lambda+1}{2p}\,P_\lambda(A/p)\,\Theta(p-|A|)\,
       C^\lambda_0(\bm{e}_p)}{n}
  \nonumber \\
  &\qquad= \sum_{\beta'\beta}V_{\beta'm}V_{\beta n}
     \sum_{\mathcal D}\sum_{(c',c)\in\mathcal P_{\mathcal D}}
     \big(\mathfrak c^{(c')}_{\beta'}\big)^{*}\mathfrak c^{(c)}_{\beta}
     \frac{2\lambda+1}{2p_\beta}\,P_\lambda(A/p_\beta)\,
     \Theta(p_\beta-|A|)
  \nonumber \\
  &\qquad\qquad\times
     \mel{\chi^{(c')}_{\beta'}}{(\sigma_0)^{\mathcal D}\,
       w^{E_M}_L(r)\,C^L_{-q}(\bm{e}_r)\,C^\lambda_0(\bm{e}_p)}
         {\chi^{(c)}_{\beta}}\,.
  \label{eq:afterbasis-EM}
\end{align}
\end{widetext}
In Eqs.~\eqref{eq:afterbasis} and \eqref{eq:afterbasis-EM} and below, $l'$ denotes $l_{c'}$, and $l'$, $j'$, and $p'=p_{\beta'}$ belong to the bra basis state $\beta'$.
The labels $l_c$, $j$, and $p=p_\beta$ belong to the ket basis state $\beta$.

\emph{(iv) Angular reduction.}
This step evaluates the last factor of Eqs.~\eqref{eq:afterbasis} and \eqref{eq:afterbasis-EM} and sums the terms of Eqs.~\eqref{eq:TI-HE} and \eqref{eq:TI-EM} over $K_n^3$ and $K_m^3$.
We write $\nu_n$ for the multiplet of the state $n$ and $\nu_m$ for the multiplet of the state $m$.
The double sum over the eigenstates in Eqs.~\eqref{eq:two-orders-HE} and \eqref{eq:two-orders-EM} is
\begin{equation}
  \sum_{n\in\occ}\ \sum_{m:\,E_m\neq E_n}
   = \sum_{\nu_n\in\occ}\ \sum_{\nu_m:\,E_m\neq E_n}\ \sum_{K_n^3K_m^3}\,,
  \label{eq:multiplets}
\end{equation}
and the prefactors and the energy denominator depend on the multiplets only.
Only the terms of Eqs.~\eqref{eq:T-HE} and \eqref{eq:T-EM} depend on $K_n^3$ and $K_m^3$, and this step sums them over the projections.
For a spherical tensor $T^\kappa_\mu$, the Wigner--Eckart theorem reads
\begin{equation}
  \begin{aligned}
  \mel{K'K^{\prime3}\,l'j'}{T^\kappa_\mu}{KK^3\,lj}
   &= \delta_{K^{\prime3},K^3+\mu}\,\mathcal W(K',\kappa,\mu;K,K^3) \\
   &\quad\times\redm{K'l'j'}{T^\kappa}{Klj}\,,
  \end{aligned}
  \label{eq:WE}
\end{equation}
with
\begin{equation}
  \begin{aligned}
  &\mathcal W(K',\kappa,\mu;K,K^3) \\
  &\quad= (-1)^{K'-K^3-\mu}
    \begin{pmatrix} K' & \kappa & K \\ -K^3-\mu & \mu & K^3 \end{pmatrix}\,.
  \end{aligned}
  \label{eq:Wdef}
\end{equation}
The reduced matrix elements of $C^\kappa$, $\sigma$, and $\tau$ between the angular states follow from the couplings in Eq.~\eqref{eq:coupled} and reduce to $3j$ and $6j$ symbols (Appendix~\ref{app:angular}).

The factor $C^\lambda_0(\bm{e}_p)$ acts on the direction of the momentum.
In position space it turns the ket component $\chi^{(c)}_\beta$ into a sum of functions $j_{l_{\mathrm p}}(p_\beta r)\, \mathcal Y_{K_{\mathrm p}K_n^3l_{\mathrm p}j_{\mathrm p}}(\bm{e}_r)$, with coefficients given by Eq.~\eqref{eq:WE} [Eq.~\eqref{eq:besselswap}].
The integral over $r$ then gives the radial integrals
\begin{equation}
  \begin{aligned}
  R^{H_E,L}_{l'l_{\mathrm p}}(p',p)
   &= \int_0^{D}\!dr\,r^2\,w^{H_E}_L(r)\,j_{l'}(p'r)\,j_{l_{\mathrm p}}(pr)\,, \\
  R^{E_M,L}_{l'l_{\mathrm p}}(p',p)
   &= \int_0^{D}\!dr\,r^2\,w^{E_M}_L(r)\,j_{l'}(p'r)\,j_{l_{\mathrm p}}(pr)\,.
  \end{aligned}
  \label{eq:radial}
\end{equation}
The remaining matrix element between angular states is reduced by inserting angular states between the operators $C^L$, $\sigma_0$, and $\tau_{-q}$ [Eq.~\eqref{eq:angchain}].
The intermediate states carry the labels $(K_{\mathrm r},j_{\mathrm r})$ and $K_{\mathrm s}$, and each operator contributes one factor $\mathcal W$ and one reduced matrix element.
The first matrix element $\mel{n}{\tau_q}{m}$ of Eqs.~\eqref{eq:T-HE} and \eqref{eq:T-EM} is reduced in the same way, with the reduced matrix element $\redm{n}{\tau}{m}$ of Eq.~\eqref{eq:redtau}.
The projections $K_n^3$ and $K_m^3$ then enter only through the Kronecker deltas of Eq.~\eqref{eq:WE} for the two isospin operators, which fix $K_m^3=K_n^3-q$, and through the factors $\mathcal W$.
Their sum over the projections defines
\begin{align}
  &g^{H_E}_q\big(K_n,K_m,K_{\mathrm p},K_{\mathrm r},K_{\mathrm s};\lambda,L\big)
  \nonumber \\
  &\quad= \sum_{K_n^3K_m^3}\delta_{K_n^3,\,K_m^3+q}\,\delta_{K_m^3,\,K_n^3-q}\,
     \mathcal W(K_n,1,q;K_m,K_m^3)
  \nonumber \\
  &\qquad\times
     \mathcal W(K_{\mathrm p},\lambda,0;K_n,K_n^3)\,
     \mathcal W(K_{\mathrm r},L,0;K_{\mathrm p},K_n^3)
  \nonumber \\
  &\qquad\times
     \big[\mathcal W(K_{\mathrm s},1,0;K_{\mathrm r},K_n^3)\big]^{[\mathcal D=1]}\,
     \mathcal W(K_m,1,-q;K_{\mathrm s},K_n^3)
  \nonumber \\
  &\quad= \sum_{K_n^3}
     \mathcal W(K_n,1,q;K_m,K_n^3{-}q)\,
     \mathcal W(K_{\mathrm p},\lambda,0;K_n,K_n^3)
  \nonumber \\
  &\qquad\times
     \mathcal W(K_{\mathrm r},L,0;K_{\mathrm p},K_n^3)\,
     \big[\mathcal W(K_{\mathrm s},1,0;K_{\mathrm r},K_n^3)\big]^{[\mathcal D=1]}
  \nonumber \\
  &\qquad\times
     \mathcal W(K_m,1,-q;K_{\mathrm s},K_n^3)\,.
  \label{eq:gsum}
\end{align}
for $H_E$ and
\begin{align}
  &g^{E_M}_q\big(K_n,K_m,K_{\mathrm p},K_{\mathrm r};\lambda,L\big)
  \nonumber \\
  &\quad= \sum_{K_n^3}
     \mathcal W(K_n,1,q;K_m,K_n^3{-}q)\,
     \mathcal W(K_{\mathrm p},\lambda,0;K_n,K_n^3)
  \nonumber \\
  &\qquad\times
     \mathcal W(K_{\mathrm r},L,-q;K_{\mathrm p},K_n^3)
  \nonumber \\
  &\qquad\times
     \big[\mathcal W(K_m,1,0;K_{\mathrm r},K_n^3{-}q)\big]^{[\mathcal D=1]}\,.
  \label{eq:gsum-EM}
\end{align}
for $E_M$.
Collecting the factors and grouping the basis states of the ket by the momentum $p_\beta$, we define
\begin{widetext}
\begin{align}
  \mathcal T^{H_E,\lambda}_{\mathrm I}(p)
  &= \sum_q(-1)^q\,\redm{n}{\tau}{m}\sum_{\beta'}\ \sum_{\beta:\,p_\beta=p}
     V_{\beta'm}V_{\beta n}
     \sum_{\mathcal D}\sum_{(c',c)\in\mathcal P_{\mathcal D}}
     \big(\mathfrak c^{(c')}_{\beta'}\big)^{*}\mathfrak c^{(c)}_{\beta}
     \sum_{L\ge0}\mathcal E^{H_E}_L
  \nonumber \\
     &\quad\times\sum_{l_{\mathrm p}}\ri^{\,l_{\mathrm p}-l_c}\,
      R^{H_E,L}_{l'l_{\mathrm p}}(p_{\beta'},p)
      \sum_{K_{\mathrm p}j_{\mathrm p}}\sum_{K_{\mathrm r}j_{\mathrm r}K_{\mathrm s}}
      \big[\delta_{K_{\mathrm s}K_{\mathrm r}}\delta_{j_{\mathrm r}j'}\big]^{[\mathcal D=0]}\,
      g^{H_E}_q\big(K_n,K_m,K_{\mathrm p},K_{\mathrm r},K_{\mathrm s};\lambda,L\big)
   \nonumber \\
   &\quad\times
      \redm{K_{\mathrm p}l_{\mathrm p}j_{\mathrm p}}{C^\lambda}{K_n l_c j}\,
      \redm{K_{\mathrm r}l'j_{\mathrm r}}{C^{L}}
           {K_{\mathrm p}l_{\mathrm p}j_{\mathrm p}}\,
      \Big[\redm{K_{\mathrm s}l'j'}{\sigma}
           {K_{\mathrm r}l'j_{\mathrm r}}\Big]^{[\mathcal D=1]}\,
      \redm{K_m l'j'}{\tau}{K_{\mathrm s}l'j'}\,,
  \label{eq:assembled} \\
  \mathcal T^{E_M,\lambda}_{\mathrm I}(p)
  &= \frac{\ri}{\sqrt2}\sum_{q=\pm1}\redm{n}{\tau}{m}\sum_{\beta'}\ \sum_{\beta:\,p_\beta=p}
     V_{\beta'm}V_{\beta n}
     \sum_{\mathcal D}\sum_{(c',c)\in\mathcal P_{\mathcal D}}
     \big(\mathfrak c^{(c')}_{\beta'}\big)^{*}\mathfrak c^{(c)}_{\beta}
     \sum_{L\ge1}\mathcal E^{E_M}_{Lq}
  \nonumber \\
     &\quad\times\sum_{l_{\mathrm p}}\ri^{\,l_{\mathrm p}-l_c}\,
      R^{E_M,L}_{l'l_{\mathrm p}}(p_{\beta'},p)
      \sum_{K_{\mathrm p}j_{\mathrm p}}\sum_{K_{\mathrm r}j_{\mathrm r}}
      \big[\delta_{K_{\mathrm r}K_m}\delta_{j_{\mathrm r}j'}\big]^{[\mathcal D=0]}\,
      g^{E_M}_q\big(K_n,K_m,K_{\mathrm p},K_{\mathrm r};\lambda,L\big)
   \nonumber \\
   &\quad\times
      \redm{K_{\mathrm p}l_{\mathrm p}j_{\mathrm p}}{C^\lambda}{K_n l_c j}\,
      \redm{K_{\mathrm r}l'j_{\mathrm r}}{C^{L}}
           {K_{\mathrm p}l_{\mathrm p}j_{\mathrm p}}\,
      \Big[\redm{K_m l'j'}{\sigma}
           {K_{\mathrm r}l'j_{\mathrm r}}\Big]^{[\mathcal D=1]}\,.
  \label{eq:assembled-EM}
\end{align}
\end{widetext}
A factor written as $[X]^{[Y]}$ equals $X$ in case $Y$ and equals $1$ otherwise.
For $\mathcal D=0$ the factor with $\sigma_0$ is absent, and the Kronecker deltas set $K_{\mathrm s}=K_{\mathrm r}$ and $j_{\mathrm r}=j'$ in Eq.~\eqref{eq:assembled} and $K_{\mathrm r}=K_m$ and $j_{\mathrm r}=j'$ in Eq.~\eqref{eq:assembled-EM}.
In Eq.~\eqref{eq:assembled-EM}, the factor $\ri/\sqrt2$ and the sum over $q$ are those of Eq.~\eqref{eq:tauy}.
Equations~\eqref{eq:assembled} and \eqref{eq:assembled-EM} with the two eigenstates exchanged give $\mathcal T^{H_E,\lambda}_{\mathrm{II}}(p)$ and $\mathcal T^{E_M,\lambda}_{\mathrm{II}}(p)$.
With these functions, the sums over $K_n^3$ and $K_m^3$ of the terms $\mathcal T^{H_E}_{X}$ of Eq.~\eqref{eq:two-orders-HE} and $\mathcal T^{E_M}_{X,q}$ of Eq.~\eqref{eq:two-orders-EM}, with $X=\mathrm I,\mathrm{II}$, are
\begin{align}
  &\sum_{K_n^3K_m^3}\mathcal T^{H_E}_{X}
  \nonumber \\
   &\quad= \sum_{\lambda}\sum_{p}\frac{2\lambda+1}{2p}\,P_\lambda(A/p)\,
     \Theta(p-|A|)\;\mathcal T^{H_E,\lambda}_{X}(p)\,,
  \nonumber \\
  &\frac{\ri}{\sqrt2\,\Delta_\perp}\sum_{q=\pm1}\sum_{K_n^3K_m^3}\mathcal T^{E_M}_{X,q}
  \nonumber \\
   &\quad= \sum_{\lambda}\sum_{p}\frac{2\lambda+1}{2p}\,P_\lambda(A/p)\,
     \Theta(p-|A|)\;\mathcal T^{E_M,\lambda}_{X}(p)\,.
  \label{eq:Tstripped}
\end{align}
For $E_M$, the left-hand side includes the factor $\ri/(\sqrt2\,\Delta_\perp)$ and the sum over $q$ of Eq.~\eqref{eq:two-orders-EM}.
In Eq.~\eqref{eq:Tstripped}, $p$ runs over the momenta of the basis states of the ket state, which is $n$ for $X=\mathrm I$ and $m$ for $X=\mathrm{II}$.
In Eq.~\eqref{eq:Tstripped}, the whole $x$ dependence is in the coefficient $(2\lambda+1)P_\lambda(A/p)\,\Theta(p-|A|)/(2p)$, which is the same for $X=\mathrm I$ and $X=\mathrm{II}$ because the operator $\hat{\mathcal O}(E_n)$ is the same in $\mathcal T_{\mathrm I}$ and $\mathcal T_{\mathrm{II}}$.

For a given $L$, $\lambda$ is restricted by a rank condition and a parity condition (Appendix~\ref{app:angular}).
The sum over $K_n^3$ vanishes unless the product of $C^\lambda$, $C^L$, and $(\sigma_0)^{\mathcal D}$ contains a spherical tensor of rank $0$ for $H_E$ and of rank $1$ for $E_M$.
The parity of the basis states requires $\lambda+L$ to be even for $\mathcal D=0$ and odd for $\mathcal D=1$.
Together the two conditions allow only
\begin{equation}
  \lambda = L\ \ (\mathcal D=0)\,,\qquad
  \lambda = L\pm1\ \ (\mathcal D=1)\,.
  \label{eq:selection}
\end{equation}
A truncation of the sum over $L$ therefore also truncates the sum over $\lambda$.

\emph{(v) Smearing of the delta function.}
This step replaces the $x$-dependent coefficient of Eq.~\eqref{eq:Tstripped} by a smooth kernel.
In a finite box the momenta $p$ are discrete.
The coefficient is then a discontinuous function of $x$, and its $x$ derivative contains delta functions.
Following Ref.~\cite{DPPPW1997} we replace the delta function of Eq.~\eqref{eq:Odef} by the Gaussian $\mathcal N_s(y) = \re^{-(y/s)^2}/(\sqrt\pi s)$ of width $s=\gamma\Mcl$.
The smearing replaces the coefficient $(2\lambda+1)P_\lambda(A/p)\Theta(p-|A|)/2p$ in Eq.~\eqref{eq:Tstripped} by the kernel
\begin{equation}
  \mathcal K_\lambda(p,x;E_n) = \frac{2\lambda+1}{2p}\int_{-p}^{p}\!dA'\,
    P_\lambda\!\left(\frac{A'}{p}\right)\mathcal N_s(A-A')\,.
  \label{eq:kernel}
\end{equation}
In Eq.~\eqref{eq:kernel}, $A$ denotes $(x+\xi)\Mcl - E_n$, as in Eq.~\eqref{eq:lcmultipole}.
The derivative term uses $\partial\mathcal K_\lambda/\partial x$.
A normalized Gaussian leaves $\int dx$ unchanged, so the smearing does not affect the sum rules \eqref{eq:sumEM} and \eqref{eq:sumHE}.
All distributions shown below are smeared, except the dash-dotted curve in Fig.~\ref{fig:decomp} (Sec.~\ref{sec:kernels}).

\emph{(vi) The form evaluated.}
This step puts the results of steps (iv) and (v) into Eqs.~\eqref{eq:two-orders-HE} and \eqref{eq:two-orders-EM}.
We decompose the sums over $n$ and $m$ as in Eq.~\eqref{eq:multiplets}, insert Eq.~\eqref{eq:Tstripped} for the sums over the projections, and replace the coefficient by the kernel of Eq.~\eqref{eq:kernel}.
This gives the two distributions in the form evaluated,
\begin{align}
  H_E(x,\xi,t) &= -\frac{\Mcl N_c}{12I}
    \sum_{\nu_n\in\occ}\ \sum_{\nu_m:\,E_m\neq E_n}\sum_{\lambda}\sum_{p}
  \nonumber \\
   &\quad\times\left[\frac{\mathcal K_\lambda(p,x;E_n)}{E_n-E_m}
     + \frac{\partial_x\mathcal K_\lambda(p,x;E_n)}{2\Mcl}\right]
  \nonumber \\
   &\quad\times\left[\mathcal T^{H_E,\lambda}_{\mathrm I}(p)
     + \mathcal T^{H_E,\lambda}_{\mathrm{II}}(p)\right]\,,
  \label{eq:final-HE} \\
  E_M(x,\xi,t) &= \frac{\ri\Mcl^2N_c}{2I}
    \sum_{\nu_n\in\occ}\ \sum_{\nu_m:\,E_m\neq E_n}\sum_{\lambda}\sum_{p}
  \nonumber \\
   &\quad\times\left[\frac{\mathcal K_\lambda(p,x;E_n)}{E_n-E_m}
     + \frac{\partial_x\mathcal K_\lambda(p,x;E_n)}{2\Mcl}\right]
  \nonumber \\
   &\quad\times\left[\mathcal T^{E_M,\lambda}_{\mathrm I}(p)
     + \mathcal T^{E_M,\lambda}_{\mathrm{II}}(p)\right]\,.
  \label{eq:final-EM}
\end{align}
The prefactors in Eqs.~\eqref{eq:final-HE} and \eqref{eq:final-EM} are the prefactors of Eqs.~\eqref{eq:oss28} and \eqref{eq:oss29}.

In the rest of this step, we omit the superscripts $H_E$ and $E_M$.
The statements hold for both distributions.
The derivative term of Eqs.~\eqref{eq:final-HE} and \eqref{eq:final-EM} contains the average of $\mathcal T_{\mathrm I}$ and $\mathcal T_{\mathrm{II}}$.
The derivative term of Eqs.~\eqref{eq:two-orders-HE} and \eqref{eq:two-orders-EM} contains $\mathcal T_{\mathrm I}$ alone.
The two forms of the derivative term are equal after the sums over $n$ and $m$.
Completeness and $[\bm\tau,\hat{\mathcal O}]=0$ give $\sum_{m=\mathrm{all}}\mathcal T_{\mathrm I} = \sum_{m=\mathrm{all}}\mathcal T_{\mathrm{II}}$ for each state $n$.
In the terms with $E_m=E_n$, both $n$ and $m$ belong to the set of states with energy $E_n$, and that set is occupied as a whole.
For such a pair, exchanging $n$ and $m$ turns $\mathcal T_{\mathrm I}$ into $\mathcal T_{\mathrm{II}}$.
Summed over that set, the terms with $E_m=E_n$ are therefore the same in the two sums.
The sums over $n\in\occ$ and $E_m\neq E_n$ are then equal as well.

In the forward limit Eqs.~\eqref{eq:final-HE} and \eqref{eq:final-EM} reduce to Eqs.~(36) and (39) of Ref.~\cite{Ossmann2005}, and $H_E(x,0,0)$ becomes the isovector unpolarized distribution $(u-d)(x)$ of Ref.~\cite{Pobylitsa1999}.

Three properties of Eqs.~\eqref{eq:final-HE} and \eqref{eq:final-EM} set the cost of an evaluation.
First, the integral over $\bm{e}_p$ and the integral over $\bm{e}_r$ are done in closed form.
The closed form follows from the expansions of step (ii) and from the orthonormality of the angular states of Appendix~\ref{app:basis}.
The radial integral \eqref{eq:radial} and the integral over $A'$ in Eq.~\eqref{eq:kernel} are the only integrals evaluated numerically.
The two integrals are one-dimensional.
Second, the kernel $\mathcal K_\lambda$ of Eq.~\eqref{eq:kernel} depends on the pair of multiplets through $E_n$ alone.
The sum over $\nu_m$ in Eqs.~\eqref{eq:final-HE} and \eqref{eq:final-EM} can therefore be taken at fixed $(\lambda,p,\nu_n)$, independently of $x$.
The sum is taken once with the factor $1/(E_n-E_m)$ and once without this factor.
Both sums are matrix products over the basis labels $\beta'$ and $\beta$ of Eqs.~\eqref{eq:assembled} and \eqref{eq:assembled-EM}.
Third, the angular coefficients and the reduced matrix elements of Eqs.~\eqref{eq:assembled} and \eqref{eq:assembled-EM} (Appendix~\ref{app:angular}) depend on the basis states only through the labels $l$ and $j$.
They depend neither on $p$ nor on $x$.
The angular algebra is therefore done once for each pair of sectors and reused for every $(p,x)$.

\section{Numerical procedure}\label{sec:numerics}

\begin{figure*}[!tbp]
\includegraphics[width=\textwidth]{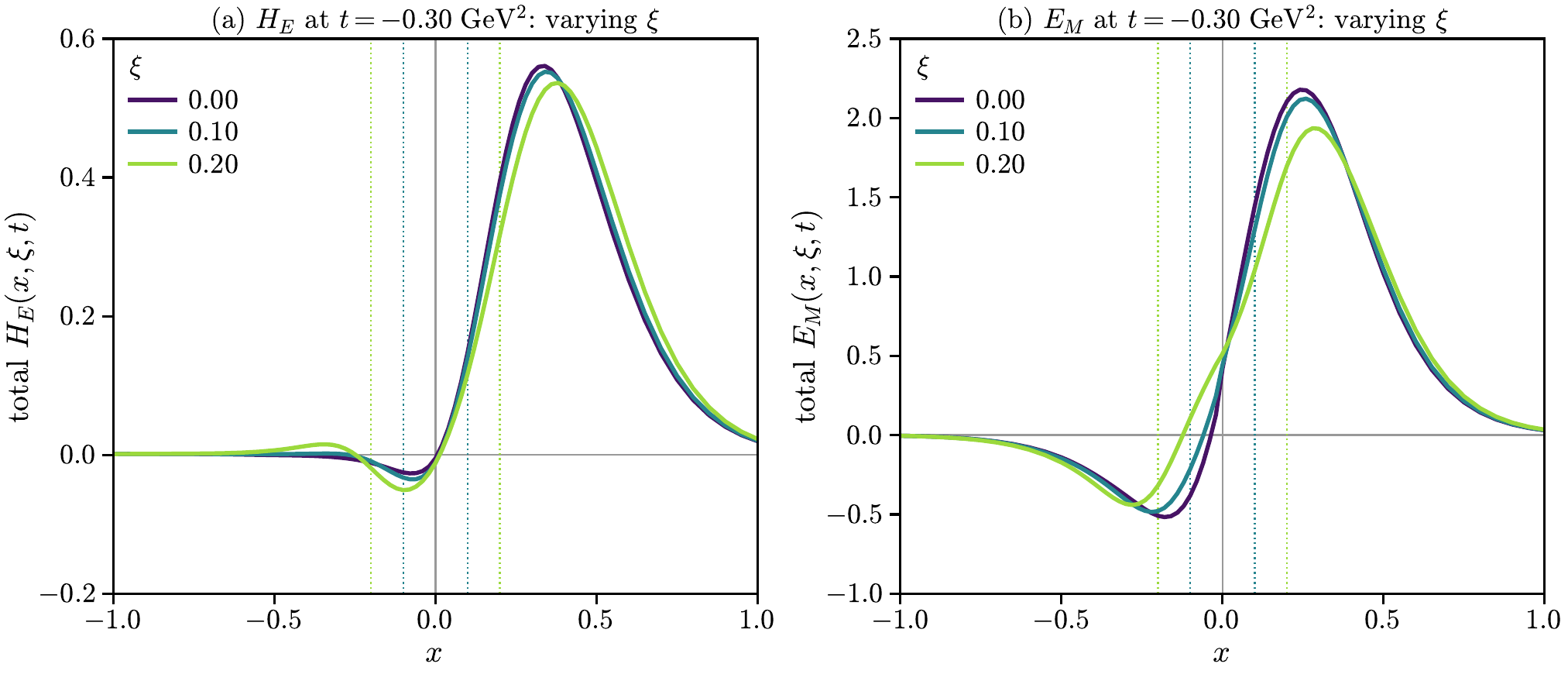}
\caption{Dependence on $\xi$ at fixed $t=-0.30$~GeV$^2$: (a) $H_E(x,\xi,t)$ and (b) $E_M(x,\xi,t)$ for $\xi=0$, $0.10$, and $0.20$, for the parameters of Sec.~\ref{sec:settings}.
The vertical dotted lines mark $x=\pm\xi$.}
\label{fig:xi-dep}
\end{figure*}

\begin{figure*}[!tbp]
\includegraphics[width=\textwidth]{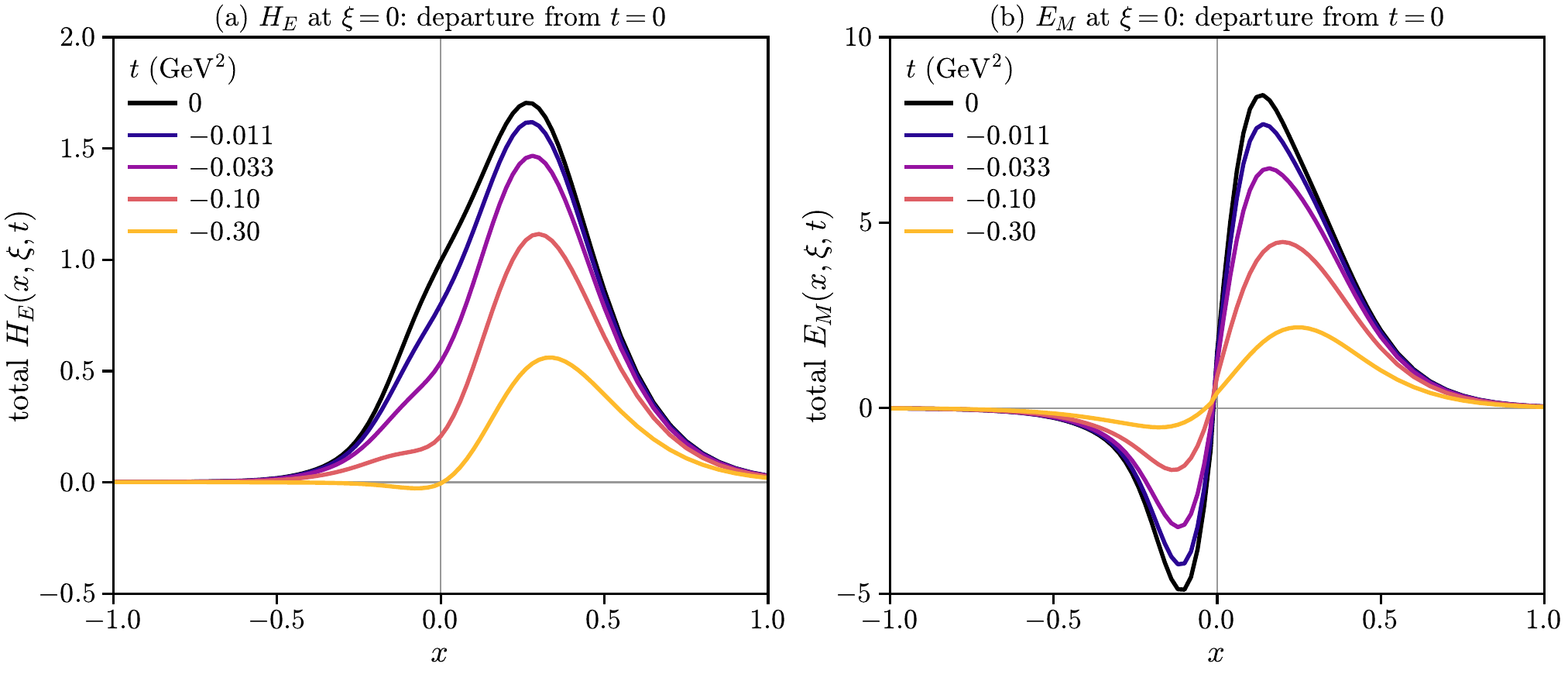}
\caption{Dependence on $t$ at $\xi=0$: (a) $H_E(x,\xi,t)$ and (b) $E_M(x,\xi,t)$ for $t=0$, $-0.011$, $-0.033$, $-0.10$, and $-0.30$~GeV$^2$.}
\label{fig:t-dep}
\end{figure*}

\begin{table}[!tbp]
\caption{First moments of the computed distributions.
The third column is $\int dx$ of the computed distribution.
The fourth column is the right-hand side $S$ of the sum rule, taken from Eq.~\eqref{eq:GEdef} for $H_E$ and from Eq.~\eqref{eq:GMdef} for $E_M$.
The right-hand side is evaluated in the basis and with the regularization of the distribution.
The fifth column is the relative difference $\left(\int dx - S\right)/S$.}
\label{tab:sumrules}
\begin{ruledtabular}
\begin{tabular}{ccccc}
$\xi$ & $t$ (GeV$^2$) & $\int dx$ & sum rule & difference \\
\hline
\multicolumn{5}{c}{$H_E$,\quad sum rule $G_E^{p-n}(t)$} \\
$0.00$ & $ 0.00$ & $0.9975$ & $1.0000$ & $-0.25\%$ \\
$0.00$ & $-0.011$ & $0.9154$ & $0.9172$ & $-0.19\%$ \\
$0.00$ & $-0.033$ & $0.7881$ & $0.7890$ & $-0.12\%$ \\
$0.00$ & $-0.10$ & $0.5544$ & $0.5544$ & $-0.01\%$ \\
$0.00$ & $-0.30$ & $0.2530$ & $0.2528$ & $+0.09\%$ \\
$0.10$ & $-0.30$ & $0.2531$ & $0.2528$ & $+0.12\%$ \\
$0.20$ & $-0.30$ & $0.2531$ & $0.2528$ & $+0.11\%$ \\
$0.10$ & $-0.50$ & $0.1307$ & $0.1305$ & $+0.21\%$ \\
$0.10$ & $-0.70$ & $0.0706$ & $0.0703$ & $+0.33\%$ \\
\hline
\multicolumn{5}{c}{$E_M$,\quad sum rule $3G_M^{p+n}(t)$} \\
$0.00$ & $ 0.00$ & $2.0253$ & $2.0286$ & $-0.17\%$ \\
$0.00$ & $-0.011$ & $1.9437$ & $1.9414$ & $+0.12\%$ \\
$0.00$ & $-0.033$ & $1.7935$ & $1.7927$ & $+0.04\%$ \\
$0.00$ & $-0.10$ & $1.4264$ & $1.4264$ & $0.00\%$ \\
$0.00$ & $-0.30$ & $0.7958$ & $0.7958$ & $0.00\%$ \\
$0.10$ & $-0.30$ & $0.7958$ & $0.7958$ & $0.00\%$ \\
$0.20$ & $-0.30$ & $0.7958$ & $0.7958$ & $+0.01\%$ \\
$0.10$ & $-0.50$ & $0.4877$ & $0.4877$ & $+0.01\%$ \\
$0.10$ & $-0.70$ & $0.3168$ & $0.3168$ & $+0.01\%$ \\
\end{tabular}
\end{ruledtabular}

\vspace{2ex}
\caption{Second moments $\int dx\,x\,E_M$ and $\int dx\,x\,H_E$ of the computed distributions.
For $E_M$ the valence and sea contributions are also listed.}
\label{tab:second-moments}
\begin{ruledtabular}
\begin{tabular}{cccccc}
 & & \multicolumn{3}{c}{$\int dx\,x\,E_M$} & $\int dx\,x\,H_E$ \\
$\xi$ & $t$ (GeV$^2$) & valence & sea & total & total \\
\hline
$0.00$ & $ 0.00$ & $0.618$ & $0.377$ & $0.995$ & $0.2400$ \\
$0.00$ & $-0.30$ & $0.281$ & $0.084$ & $0.365$ & $0.1055$ \\
$0.10$ & $-0.30$ & $0.281$ & $0.084$ & $0.365$ & $0.1069$ \\
$0.20$ & $-0.30$ & $0.281$ & $0.085$ & $0.365$ & $0.1111$ \\
$0.10$ & $-0.50$ & $0.184$ & $0.044$ & $0.228$ & $0.0679$ \\
$0.10$ & $-0.70$ & $0.126$ & $0.024$ & $0.150$ & $0.0449$ \\
\end{tabular}
\end{ruledtabular}
\end{table}

\subsection{Settings}\label{sec:settings}

We work in the chiral limit with $M=350$~MeV, $f_\pi=93$~MeV and $\expval{\bar qq} = -(286.5~\mathrm{MeV})^3$.
For these values, Eqs.~\eqref{eq:pvconditions}--\eqref{eq:condensate-condition} give $\Lambda_1=634.05$~MeV, $c_1=+0.35039$, $\Lambda_2=1505.53$~MeV, and $c_2=-0.0081019$.

The Kahana--Ripka basis~\cite{KahanaRipka} of Sec.~\ref{sec:reduction} is set in a spherical box of radius $D$.
In each $(K,\Pi)$ sector the momenta are discretized by $j_K(p_iD)=0$ and kept up to a momentum $k_{\max}$, and $H_c$ is diagonalized in the sector.
We use $D = 30/M$, and we raise $k_{\max}$ until the computed distributions no longer change.
The basis states, their coefficients \eqref{eq:coeffs}, and their energies (Appendix~\ref{app:basis}) are rebuilt at each of the three quark masses.

The self-consistent profile is computed once in the same box.
Beyond a radius $r_A$ we replace the tail of $F(r)$ by the Yukawa form
\begin{equation}
    F(r) = F(r_A)\left(\frac{r_A}{r}\right)^{2}
    \re^{-m_{\mathrm{t}}(r-r_A)}\,
    \frac{1+m_{\mathrm{t}}r}{1+m_{\mathrm{t}}r_A}\,,
  \label{eq:patch}
\end{equation}
where $m_{\mathrm t}$ is a mass that suppresses the tail.
We take $r_A = 4$~fm and $m_{\mathrm t}=200$~MeV from Ref.~\cite{Ossmann2005}.
Without the replacement, the sums over the occupied states and over the non-occupied states give different distributions in the box~\cite{Ossmann2005}.
The resulting soliton has $\Mcl = 1061.8$~MeV and $I = 7.694\times10^{-3}$~MeV$^{-1}$.

The same six kinematic points are used for both distributions, $(\xi,t) = (0,0)$, $(0,-0.30)$, $(0.10,-0.30)$, $(0.20,-0.30)$, $(0.10,-0.50)$, and $(0.10,-0.70)$, with $t$ in GeV$^2$.
All six points satisfy $-t\ge4\Mcl^2\xi^2$.
Three further points, at $\xi=0$ and $t=-0.011$, $-0.033$, and $-0.10$~GeV$^2$, show how the distributions depart from the forward limit.

\subsection{Kernels and truncations}\label{sec:kernels}

The results below use $\gamma=0.10$, and the sums over $L$ are truncated at $L_{\max}=16$.
Appendix~\ref{app:lmax} justifies this truncation.
The isospin matrix element $\mel{n}{\tau^a}{m}$ vanishes unless the two states have the same parity and $|K_n-K_m|\le1$.
The double sum therefore runs over the same pairs of sectors as at $t=0$.
Each triple $(L,\lambda,\mathcal D)$ allowed by Eq.~\eqref{eq:selection} requires a separate angular coefficient and a separate radial integral.
The cost of a kinematic point is therefore proportional to the number of allowed triples with $L\le L_{\max}$.

The model distributions are functions of $x\Mcl$.
In the large-$N_c$ limit $\Mcl$ is of order $N_c$, and the range $|x\Mcl|\le\Mcl$ extends to $-\infty<x\Mcl<\infty$.
The sum rules \eqref{eq:sumEM} and \eqref{eq:sumHE} follow from the integral over this range~\cite{Ossmann2005}.
For the finite value of $\Mcl$ the distributions do not vanish for $|x|>1$, and the integral includes this region.
We evaluate the integrals over $|x|\le2$.
At $|x|=2$ the magnitude of every computed distribution is below $2\times10^{-5}$.
The smearing turns a pole of a distribution into a finite peak and a finite dip.
In the chiral limit, the sea contribution to $E_M$ at $\xi=0$ and $t=0$ has a pole at $x=0$~\cite{Ossmann2005}.
In the leading order of the expansion of the sea contribution in gradients of the pion field, we find no pole for $\bm\Delta\neq0$.
We therefore remove the smearing only from the sea contribution to $E_M$ at $\xi=0$ and $t=0$, with the procedure of Ref.~\cite{DPPPW1997}.

\section{Results}\label{sec:results}

\begin{figure*}[!tbp]
\includegraphics[width=\textwidth]{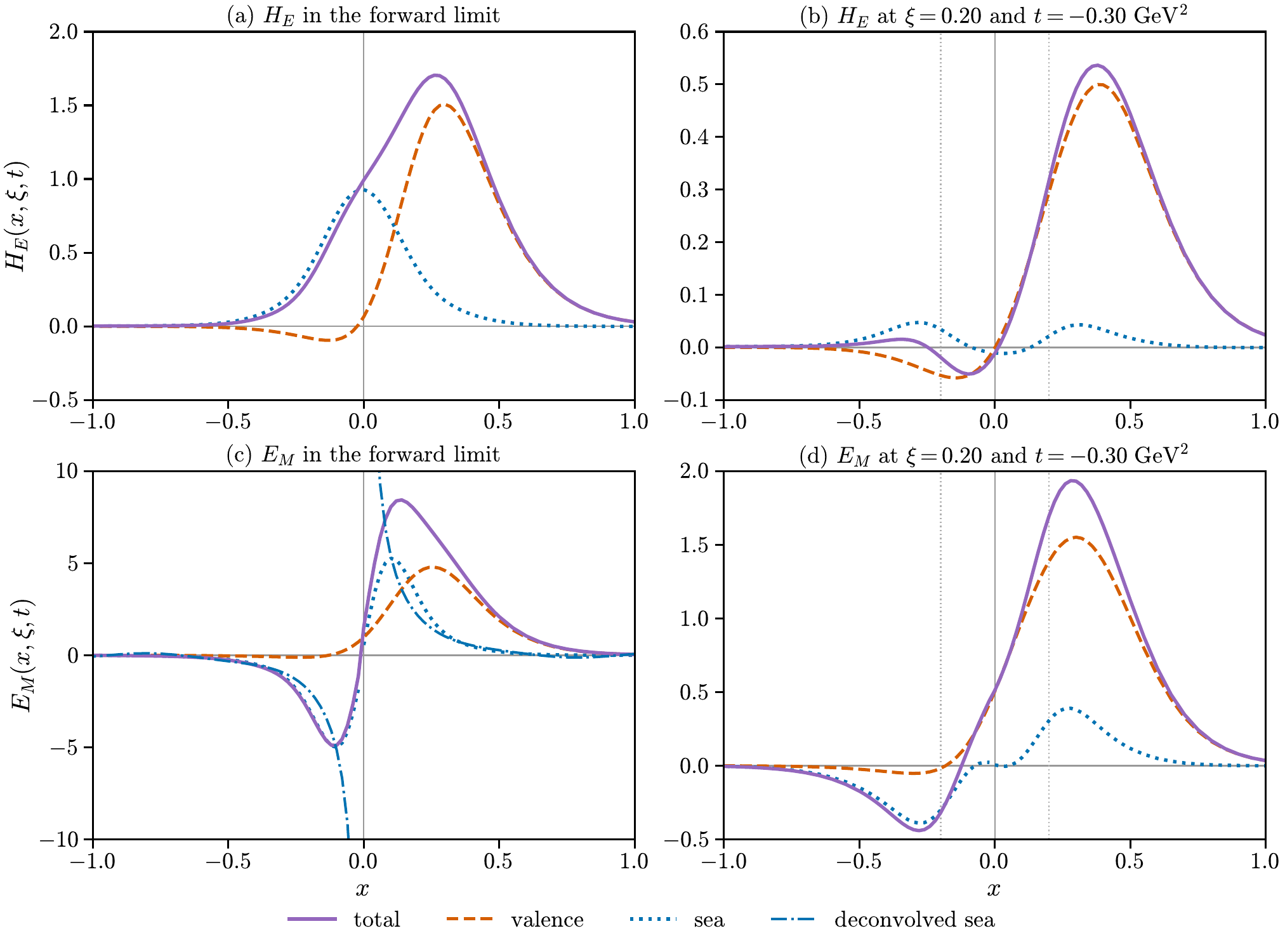}
\caption{Valence contribution (dashed) and sea contribution (dotted) to $H_E(x,\xi,t)$ [(a) and (b)] and to $E_M(x,\xi,t)$ [(c) and (d)], with their sum (solid).
The left column is the forward limit $\xi=0$, $t=0$, and the right column is $\xi=0.20$, $t=-0.30$~GeV$^2$.
In panel (c) the dash-dotted curve is the sea contribution with the smearing removed (Sec.~\ref{sec:kernels}).
The vertical dotted lines mark $x=\pm\xi$.}
\label{fig:decomp}
\end{figure*}

For $x\ge0$ all curves are computed with the sums over the occupied states of Sec.~\ref{sec:spectral}, and for $x<0$ they are computed with the sums over the non-occupied states~\cite{Pobylitsa1999}.
Table~\ref{tab:sumrules} compares the first moment of each computed distribution with the right-hand side of Eq.~\eqref{eq:sumHE} or Eq.~\eqref{eq:sumEM}.
The two sides agree to $0.33\%$ or better at all eighteen points.
The two sets of sums give different curves mainly for $|x|<0.5$.

Figures~\ref{fig:xi-dep}--\ref{fig:decomp} show the $x$ dependence for $-t\le0.30$~GeV$^2$.
The two points at $\xi=0.10$ with $t=-0.50$ and $-0.70$~GeV$^2$ enter only Tables~\ref{tab:sumrules} and \ref{tab:second-moments}.

\subsection{Dependence on \texorpdfstring{$\xi$}{xi}}\label{sec:xi-dependence}

Figure~\ref{fig:xi-dep} shows $H_E$ and $E_M$ at $t=-0.30$~GeV$^2$ for $\xi=0$, $0.10$, and $0.20$.
For both distributions the dependence on $\xi$ is weak compared with the dependence on $t$ of Sec.~\ref{sec:t-dependence}.
The curves differ mainly for $|x|<0.5$.
The negative minimum of $H_E$ near $x=-0.1$ deepens from $-0.03$ at $\xi=0$ to $-0.05$ at $\xi=0.20$.

\subsection{Dependence on \texorpdfstring{$t$}{t}}\label{sec:t-dependence}

Figure~\ref{fig:t-dep} shows $H_E$ and $E_M$ at $\xi=0$ for $t$ from $0$ to $-0.30$~GeV$^2$.
For both distributions the dependence on $t$ is strong.
At $t=-0.30$~GeV$^2$, $H_E$ has a negative minimum near $x=-0.1$, which is absent for $-t\le0.10$~GeV$^2$.
The maximum and the minimum of $E_M$ shrink and move away from $x=0$ as $-t$ grows.

\subsection{Valence and sea contributions}\label{sec:decomposition}

Figure~\ref{fig:decomp} shows the valence and sea contributions to $H_E$ and $E_M$ in the forward limit and at $\xi=0.20$, $t=-0.30$~GeV$^2$.
In the forward limit the sea contribution to $H_E$ is comparable to the valence contribution, and the sea contribution to $E_M$ has a pole at $x=0$ (Sec.~\ref{sec:kernels}).
The dash-dotted curve in panel (c) is the sea contribution with the smearing removed.
For $t<0$ the sea contribution is bounded, and the valence contribution dominates the maximum of both distributions.
Panels (b) and (d) show this at $\xi=0.20$ and $t=-0.30$~GeV$^2$.
There, for $H_E$ at $x<0$ the valence contribution is negative and the sea contribution is positive.
Their sum gives the negative minimum near $x=-0.1$ and the small positive bump near $x=-0.3$ of Fig.~\ref{fig:xi-dep}(a).
The negative part of $E_M$ at $x<0$ comes mainly from the sea contribution.
The sea contribution to $E_M$ is small near $x=0$ and grows toward $|x|=\xi$: its magnitude is below $0.03$ for $|x|\le0.08$ and reaches $0.30$ at $|x|=0.20$.

\subsection{Moments}\label{sec:second-moments}

At $\xi=0$ and $t=0$, Table~\ref{tab:sumrules} gives $\int dx\,E_M = 2.025$.
This value corresponds to the isoscalar magnetic moment $\mu^{p+n} = \tfrac13\int dx\,E_M = 0.68$ in nuclear magnetons.
Table~\ref{tab:second-moments} lists the second moments of $E_M$ and of $H_E$.
The second moment of $E_M$ does not depend on $\xi$, because the terms of order $\xi^2$ of $H^{u+d}$ and of $E^{u+d}$ cancel in $E_M$~\cite{Ossmann2005}.
At $t=-0.30$~GeV$^2$ the computed values at the three values of $\xi$ agree to $0.13\%$.
Polynomiality~\cite{Ji1998JPG} restricts the second moment of $H_E$ to the form $a+b\xi^2$~\cite{Ossmann2005}.
At $t=-0.30$~GeV$^2$ the computed values at $\xi=0$ and at $\xi=0.10$ give $a=0.1055$ and $b=0.140$.
The form then gives $0.1111$ at $\xi=0.20$, which is the computed value.
For the second moment of $E_M$, the sea contribution divided by the valence contribution falls from $0.61$ at $t=0$ to $0.19$ at $t=-0.70$~GeV$^2$.

\section{Discussion}\label{sec:discussion}

Section~\ref{sec:results} presents the $x$ dependence of $E_M$ and $H_E$ in the CQSM at nonzero $\xi$ and $t$.
The contribution of the Dirac continuum is summed over the quark states.
To our knowledge, no previous result for an $x$ dependence at nonzero $\xi$ and $t$ contains this sum.
The results are also the first for the unpolarized subleading combinations at nonzero $\xi$ and $t$.
The calculation uses no ansatz for the dependence on $\xi$ and $t$, no interpolation formula, and no gradient expansion.
The calculation uses $m_\pi=0$ and the large-$N_c$ kinematics of Sec.~\ref{sec:kinematics}, which restricts the model to $|t|\ll M_N^2$~\cite{Ossmann2005, Goeke2007}.

The model expressions have exact properties that hold only when the sums run over all quark states.
Because all quark states are summed, the results can be tested numerically against these properties.
The results pass five numerical tests.
First, each first moment agrees with the corresponding form factor (Table~\ref{tab:sumrules}).
The form factor is computed as a separate double sum over the same quark states.
Second, the first moments do not depend on $\xi$, as the form factors require.
Third, the second moment of $E_M$ does not depend on $\xi$ (Table~\ref{tab:second-moments}).
Fourth, the second moment of $H_E$ has the form $a+b\xi^2$ required by polynomiality, where $a=0.1055$ and $b=0.140$.
Fifth, the second moment of $E_M$ at $t=0$ is $2J^{u+d}(0)$~\cite{Ji1997}, which equals $1$ in the CQSM because the quarks carry the whole angular momentum of the nucleon~\cite{Ossmann2005}.
The computed value is $0.995$.
The results are also even in $\xi$, as the model expressions require~\cite{Ossmann2005}.

The results can also be compared with the previous calculation.
At $\xi=0$ and $t=0$ the first moment of $E_M$ can be compared with Ref.~\cite{Ossmann2005}.
The isoscalar magnetic moment of Sec.~\ref{sec:results} is $0.68$, and Ref.~\cite{Ossmann2005} gives $0.65$.
Reference~\cite{Ossmann2005} uses a single Pauli--Villars subtraction.

The sea contribution decreases with $-t$ faster than the valence contribution.
For $H_E$ and for $E_M$ this is seen by comparing the forward limit with the point $\xi=0.20$, $t=-0.30$~GeV$^2$ in Fig.~\ref{fig:decomp}.
For the second moment of $E_M$ it is seen at every computed $t$ in Table~\ref{tab:second-moments}.
The Dirac continuum describes the pion cloud of the soliton, which extends to larger distances than the valence state~\cite{Christov:1995vm}.
A faster decrease with $-t$ is therefore expected.

The reduction of Sec.~\ref{sec:reduction} applies to any Dirac matrix and any isospin matrix in the light-cone operator.
The light-cone operators of the other GPDs of the CQSM differ from the operators of Eqs.~\eqref{eq:oss28} and \eqref{eq:oss29} only in these matrices and in $c$-number factors~\cite{Petrov1998, Penttinen2000}.
We therefore expect that the other GPDs can be computed in the same way.
With $H^{u+d}$ and $E^{u-d}$ at the kinematic points of this paper, $E_M$ and $H_E$ would give $E^{u+d}$ and $H^{u-d}$ separately.
The subleading combinations $\tilde H^{u+d}$ and $\tilde E^{u+d}$ and the chiral-odd GPDs~\cite{Wakamatsu2009, KimWeiss2025, Kim2025} can be treated in the same way.
Reference~\cite{Kim2025} states that a complete prediction of the subleading chiral-odd GPDs requires the sum over all quark states.
The present calculation shows that this sum can be carried out at nonzero $\xi$ and $t$.

\begin{acknowledgments}
The author thanks Kenji Fukushima for his guidance and Tomoya Uji for useful discussions.
The author used Anthropic's Claude to assist with the implementation and debugging of the numerical code and in the preparation of the manuscript.
This work was supported by IIW, WINGS Program, The University of Tokyo.
\end{acknowledgments}

\appendix

\section{Kahana--Ripka basis}\label{app:basis}

This appendix gives the details of the basis of step (iii) of Sec.~\ref{sec:reduction}.
The basis states $\ket{\beta K^3}$ of Eq.~\eqref{eq:expansion} are the solutions of Eq.~\eqref{eq:dirac-spectrum} at $F\equiv0$ in the box.
The Hamiltonian $H_c$ is block diagonal in the sectors $(K,\Pi)$.
Within a sector $(K_\beta,\Pi_\beta)$, a basis state is specified by $j=K_\beta\pm\tfrac12$, by the sign of the energy $\varepsilon_\beta=\pm\sqrt{p_\beta^2+M^2}$, by the discrete momentum $p_\beta$, and by the projection $K^3$.
We write $\beta$ for the sector and the first three labels.

The two components of Eq.~\eqref{eq:coupled} have the same $j$, and their orbital angular momenta are
\begin{equation}
  \begin{aligned}
  \Pi_\beta = (-1)^{K_\beta}\ &:\ (l_{\mathrm{up}},l_{\mathrm{dn}})
    = (K_\beta,\,K_\beta\pm1)\,, \\
  \Pi_\beta = (-1)^{K_\beta+1}\ &:\ (l_{\mathrm{up}},l_{\mathrm{dn}})
    = (K_\beta\pm1,\,K_\beta)\,.
  \end{aligned}
  \label{eq:channels}
\end{equation}
The upper sign holds for $j=K_\beta+\tfrac12$, and the lower sign holds for $j=K_\beta-\tfrac12$.
The basis states with $j=K_\beta-\tfrac12$ do not exist for $K_\beta=0$.
In all four cases the basis state has parity $(-1)^{l_{\mathrm{up}}}$.
All signs in the Racah algebra below assume the coupling order of Eq.~\eqref{eq:coupled}, $\hat{\bm K} = (\hat{\bm L}+\hat{\bm S})+\hat{\bm T}$.

A matrix element between two angular states $\ket{KK^3\,lj}$ is an integral over $\bm{e}_r$ alone.

The coefficients $\mathfrak c^{(c)}_\beta$ are those of the free Dirac spinor in the box.
Explicitly, the coefficients are
\begin{equation}
  \begin{aligned}
  \big(\mathfrak c^{(\mathrm{up})}_\beta,\mathfrak c^{(\mathrm{dn})}_\beta\big)
   &= N_\beta\big(\ri,\,-s_\beta\alpha_\beta\big)\,,
   && \varepsilon_\beta>0\,, \\
  \big(\mathfrak c^{(\mathrm{up})}_\beta,\mathfrak c^{(\mathrm{dn})}_\beta\big)
   &= N_\beta\big(\ri\alpha_\beta,\,s_\beta\big)\,,
   && \varepsilon_\beta<0\,,
  \end{aligned}
  \label{eq:coeffs}
\end{equation}
where $\alpha_\beta$ denotes $p_\beta/(|\varepsilon_\beta|+M)$~\cite{WakamatsuYoshiki}.
The sign $s_\beta$ is $-1$ for $j=l_{\mathrm{up}}+\tfrac12$ and $+1$ for $j=l_{\mathrm{up}}-\tfrac12$.
The overall phase is fixed by the requirement that the matrix of $H_c$ in the basis and the coefficients $V$ are real.
The constant $N_\beta>0$ follows from the normalization $\braket{\beta}{\beta}=1$,
\begin{equation}
  \begin{aligned}
  &\int_0^{D}\!dr\,r^2\Big[\big|\mathfrak c^{(\mathrm{up})}_\beta\big|^2
     j_{l_{\mathrm{up}}}(p_\beta r)^2
   + \big|\mathfrak c^{(\mathrm{dn})}_\beta\big|^2
     j_{l_{\mathrm{dn}}}(p_\beta r)^2\Big] = 1\,, \\
  &\int_0^{D}\!dr\,r^2 j_l(pr)^2
   = \frac{D^3}{2}\Big[j_l(pD)^2-j_{l-1}(pD)\,j_{l+1}(pD)\Big]\,.
  \end{aligned}
  \label{eq:norm}
\end{equation}
Here, \(D\) denotes the radius of the box.

Inserting Eqs.~\eqref{eq:expansion}, \eqref{eq:coupled}, and \eqref{eq:dirac-split} into the second matrix element of Eqs.~\eqref{eq:TI-HE} and \eqref{eq:TI-EM} gives Eqs.~\eqref{eq:afterbasis} and \eqref{eq:afterbasis-EM} of step (iii) of Sec.~\ref{sec:reduction}.

\section{Angular reduction}\label{app:angular}

This appendix evaluates the last factor of Eqs.~\eqref{eq:afterbasis} and \eqref{eq:afterbasis-EM} and sums the terms of Eqs.~\eqref{eq:TI-HE} and \eqref{eq:TI-EM} over $K_n^3$ and $K_m^3$.
The result is Eqs.~\eqref{eq:assembled} and \eqref{eq:assembled-EM} of step (iv) of Sec.~\ref{sec:reduction}, and the rank and parity conditions at the end of this appendix give the selection rule \eqref{eq:selection}.

The reduced matrix elements of the Wigner--Eckart theorem \eqref{eq:WE} follow from the two couplings in Eq.~\eqref{eq:coupled}, from the reduction formulas for coupled states~\cite{Edmonds, Varshalovich}, and from
\begin{equation}
  \begin{aligned}
  \redm{\tfrac12_{\mathrm{spin}}}{\sigma}{\tfrac12_{\mathrm{spin}}}
   &= \redm{\tfrac12_{\mathrm{iso}}}{\tau}{\tfrac12_{\mathrm{iso}}}
    = \sqrt6\,, \\
  \redm{l'}{C^\kappa}{l} &= (-1)^{l'}\sqrt{(2l'{+}1)(2l{+}1)}
    \begin{pmatrix} l' & \kappa & l \\ 0 & 0 & 0 \end{pmatrix}\,.
  \end{aligned}
  \label{eq:elementary}
\end{equation}
Here $\tfrac12_{\mathrm{spin}}$ denotes the spin of the quark, and $\tfrac12_{\mathrm{iso}}$ denotes the isospin of the quark.

The factor $C^\lambda_0(\bm{e}_p)$ acts on the direction of the momentum.
We reduce the product $C^\lambda_0(\bm{e}_p)\,\mathcal Y_{K_\beta K^3l_cj}(\bm{e}_p)$ in momentum space with Eq.~\eqref{eq:WE}.
In position space, the result is
\begin{equation}
  \begin{aligned}
  \mel{\bm r}{C^\lambda_0(\bm{e}_p)}{\chi^{(c)}_\beta}
   &= \ri^{-l_c}\!\int\!\frac{d^3p}{(2\pi)^3}\,
      \frac{2\pi^2}{p_\beta^2}\,\delta(|\bm p|-p_\beta)\,
      \re^{\ri\bm p\cdot\bm r} \\
   &\quad\times C^\lambda_0(\bm{e}_p)\,
      \mathcal Y_{K_\beta K^3l_cj}(\bm{e}_p) \\
   &= \sum_{K_{\mathrm p}l_{\mathrm p}j_{\mathrm p}}
      \ri^{\,l_{\mathrm p}-l_c}\,
      \mathcal W(K_{\mathrm p},\lambda,0;K_\beta,K^3) \\
   &\quad\times
      \redm{K_{\mathrm p}l_{\mathrm p}j_{\mathrm p}}{C^\lambda}{K_\beta l_c j}\,
      j_{l_{\mathrm p}}(p_\beta r) \\
   &\quad\times
      \mathcal Y_{K_{\mathrm p}K^3l_{\mathrm p}j_{\mathrm p}}(\bm{e}_r)\,.
  \end{aligned}
  \label{eq:besselswap}
\end{equation}
In Eq.~\eqref{eq:besselswap}, $K_{\mathrm p}$, $l_{\mathrm p}$, and $j_{\mathrm p}$ label the states in which $C^\lambda_0(\bm{e}_p)\,\chi^{(c)}_\beta$ is expanded.

For the ket basis states in Eqs.~\eqref{eq:afterbasis} and \eqref{eq:afterbasis-EM}, $K_\beta=K_n$ and $K^3=K_n^3$.
Putting Eq.~\eqref{eq:besselswap} into the last factor of these equations separates the factor into a radial integral and an angular matrix element,
\begin{widetext}
\begin{align}
  &\mel{\chi^{(c')}_{\beta'}}{\tau_{-q}\,(\sigma_0)^{\mathcal D}\,
      w^{H_E}_L(r)\,C^L_0(\bm{e}_r)\,C^\lambda_0(\bm{e}_p)}
      {\chi^{(c)}_{\beta}}
  \nonumber \\
  &\qquad= \sum_{K_{\mathrm p}l_{\mathrm p}j_{\mathrm p}}
     \ri^{\,l_{\mathrm p}-l_c}\,
     \mathcal W(K_{\mathrm p},\lambda,0;K_n,K_n^3)\,
     \redm{K_{\mathrm p}l_{\mathrm p}j_{\mathrm p}}{C^\lambda}{K_n l_c j}\;
     R^{H_E,L}_{l'l_{\mathrm p}}(p_{\beta'},p_\beta)
  \nonumber \\
  &\qquad\qquad\times
     \mel{K_m K_m^3\,l'j'}{\tau_{-q}\,(\sigma_0)^{\mathcal D}\,
       C^L_0(\bm{e}_r)}{K_{\mathrm p}K_n^3\,l_{\mathrm p}j_{\mathrm p}}\,,
  \label{eq:factorized} \\
  &\mel{\chi^{(c')}_{\beta'}}{(\sigma_0)^{\mathcal D}\,
      w^{E_M}_L(r)\,C^L_{-q}(\bm{e}_r)\,C^\lambda_0(\bm{e}_p)}
      {\chi^{(c)}_{\beta}}
  \nonumber \\
  &\qquad= \sum_{K_{\mathrm p}l_{\mathrm p}j_{\mathrm p}}
     \ri^{\,l_{\mathrm p}-l_c}\,
     \mathcal W(K_{\mathrm p},\lambda,0;K_n,K_n^3)\,
     \redm{K_{\mathrm p}l_{\mathrm p}j_{\mathrm p}}{C^\lambda}{K_n l_c j}\;
     R^{E_M,L}_{l'l_{\mathrm p}}(p_{\beta'},p_\beta)
  \nonumber \\
  &\qquad\qquad\times
     \mel{K_m K_m^3\,l'j'}{(\sigma_0)^{\mathcal D}\,
       C^L_{-q}(\bm{e}_r)}{K_{\mathrm p}K_n^3\,l_{\mathrm p}j_{\mathrm p}}\,.
  \label{eq:factorized-EM}
\end{align}
\end{widetext}
The radial integrals $R$ are those of Eq.~\eqref{eq:radial}.
In Eqs.~\eqref{eq:factorized} and \eqref{eq:factorized-EM} and below, $l'$, $j'$, and $p'$ belong to the bra basis state.
The labels $l_c$, $j$, and $p$ belong to the ket basis state.
We now evaluate the last factor of Eqs.~\eqref{eq:factorized} and \eqref{eq:factorized-EM}.
This factor contains three operators for $H_E$ with $\mathcal D=1$ and fewer operators in the other cases.
We insert states $\ket{KK^3\,lj}$ between adjacent operators and reduce each resulting matrix element with Eq.~\eqref{eq:WE}.
The insertion and the reduction give
\begin{widetext}
\begin{align}
  &\mel{K_m K_m^3\,l'j'}{\tau_{-q}\,(\sigma_0)^{\mathcal D}\,
       C^L_0(\bm{e}_r)}{K_{\mathrm p}K_n^3\,l_{\mathrm p}j_{\mathrm p}}
  \nonumber \\
  &\qquad= \sum_{K_{\mathrm r}j_{\mathrm r}K_{\mathrm s}}
     \big[\delta_{K_{\mathrm s}K_{\mathrm r}}\delta_{j_{\mathrm r}j'}\big]^{[\mathcal D=0]}\,
     \mel{K_m K_m^3\,l'j'}{\tau_{-q}}{K_{\mathrm s}K_n^3\,l'j'}\,
     \Big[\mel{K_{\mathrm s}K_n^3\,l'j'}{\sigma_0}
          {K_{\mathrm r}K_n^3\,l'j_{\mathrm r}}\Big]^{[\mathcal D=1]}
  \nonumber \\
  &\qquad\qquad\times
     \mel{K_{\mathrm r}K_n^3\,l'j_{\mathrm r}}{C^L_0(\bm{e}_r)}
         {K_{\mathrm p}K_n^3\,l_{\mathrm p}j_{\mathrm p}}
  \nonumber \\
  &\qquad= \delta_{K_m^3,\,K_n^3-q}\sum_{K_{\mathrm r}j_{\mathrm r}K_{\mathrm s}}
     \big[\delta_{K_{\mathrm s}K_{\mathrm r}}\delta_{j_{\mathrm r}j'}\big]^{[\mathcal D=0]}\,
     \mathcal W(K_m,1,-q;K_{\mathrm s},K_n^3)\,
     \redm{K_m l'j'}{\tau}{K_{\mathrm s}l'j'}
  \nonumber \\
  &\qquad\qquad\times
     \Big[\mathcal W(K_{\mathrm s},1,0;K_{\mathrm r},K_n^3)\,
     \redm{K_{\mathrm s}l'j'}{\sigma}{K_{\mathrm r}l'j_{\mathrm r}}
     \Big]^{[\mathcal D=1]}\,
     \mathcal W(K_{\mathrm r},L,0;K_{\mathrm p},K_n^3)\,
     \redm{K_{\mathrm r}l'j_{\mathrm r}}{C^{L}}
          {K_{\mathrm p}l_{\mathrm p}j_{\mathrm p}}\,,
  \label{eq:angchain} \\
  &\mel{K_m K_m^3\,l'j'}{(\sigma_0)^{\mathcal D}\,
       C^L_{-q}(\bm{e}_r)}{K_{\mathrm p}K_n^3\,l_{\mathrm p}j_{\mathrm p}}
  \nonumber \\
  &\qquad= \delta_{K_m^3,\,K_n^3-q}\sum_{K_{\mathrm r}j_{\mathrm r}}
     \big[\delta_{K_{\mathrm r}K_m}\delta_{j_{\mathrm r}j'}\big]^{[\mathcal D=0]}\,
     \Big[\mathcal W(K_m,1,0;K_{\mathrm r},K_n^3{-}q)\,
     \redm{K_m l'j'}{\sigma}{K_{\mathrm r}l'j_{\mathrm r}}
     \Big]^{[\mathcal D=1]}
  \nonumber \\
  &\qquad\qquad\times
     \mathcal W(K_{\mathrm r},L,-q;K_{\mathrm p},K_n^3)\,
     \redm{K_{\mathrm r}l'j_{\mathrm r}}{C^{L}}
          {K_{\mathrm p}l_{\mathrm p}j_{\mathrm p}}\,.
  \label{eq:angchain-EM}
\end{align}
\end{widetext}
A factor written as $[X]^{[Y]}$ equals $X$ in case $Y$ and equals $1$ otherwise.
The Kronecker delta $\delta_{K_m^3,\,K_n^3-q}$ comes from Eq.~\eqref{eq:WE} for the matrix element with the bra $\bra{K_mK_m^3\,l'j'}$.
For $\mathcal D=0$ the factor with $\sigma_0$ is absent.
The Kronecker deltas then set $K_{\mathrm s}=K_{\mathrm r}$ and $j_{\mathrm r}=j'$ in Eq.~\eqref{eq:angchain}, and they set $K_{\mathrm r}=K_m$ and $j_{\mathrm r}=j'$ in Eq.~\eqref{eq:angchain-EM}.

The first matrix element in Eqs.~\eqref{eq:T-HE} and \eqref{eq:T-EM} is reduced with Eq.~\eqref{eq:WE},
\begin{equation}
  \begin{aligned}
  \mel{n}{\tau_q}{m} &= \delta_{K_n^3,\,K_m^3+q}\,
     \mathcal W(K_n,1,q;K_m,K_m^3) \\
   &\quad\times\redm{n}{\tau}{m}\,.
  \end{aligned}
  \label{eq:firstme}
\end{equation}
Because $\tau_q$ is even under parity, the matrix element of Eq.~\eqref{eq:firstme} vanishes unless $n$ and $m$ have the same parity.
The reduced matrix element follows from the basis expansion,
\begin{equation}
  \begin{aligned}
  \redm{n}{\tau}{m} &= \sum_{\beta\beta'}V_{\beta n}V_{\beta'm}
     \sum_{c}\big(\mathfrak c^{(c)}_{\beta}\big)^{*}\mathfrak c^{(c)}_{\beta'}\,
     \delta_{l_cl_c'}\,\delta_{jj'} \\
   &\quad\times\redm{K_nl_cj}{\tau}{K_ml_cj} \\
   &\quad\times
     \int_0^{D}\!dr\,r^2\,j_{l_c}(p_\beta r)\,j_{l_c}(p_{\beta'}r)\,.
  \end{aligned}
  \label{eq:redtau}
\end{equation}
In Eq.~\eqref{eq:redtau}, $l_c$ and $j$ belong to the basis state $\beta$, and $l_c'$ and $j'$ belong to the basis state $\beta'$.

In the product of Eqs.~\eqref{eq:factorized}, \eqref{eq:angchain}, and \eqref{eq:firstme}, $K_n^3$ and $K_m^3$ enter only through the Kronecker deltas and the factors $\mathcal W$.
Their sum over $K_n^3$ and $K_m^3$ is the coefficient $g^{H_E}_q$ of Eq.~\eqref{eq:gsum}.
The first Kronecker delta of Eq.~\eqref{eq:gsum} comes from Eq.~\eqref{eq:firstme}, and the second comes from Eq.~\eqref{eq:angchain}.
The two impose the same condition, which gives the second equality of Eq.~\eqref{eq:gsum}.
For $E_M$, the product of Eqs.~\eqref{eq:factorized-EM}, \eqref{eq:angchain-EM}, and \eqref{eq:firstme} gives Eq.~\eqref{eq:gsum-EM} in the same way.
Collecting Eqs.~\eqref{eq:afterbasis}--\eqref{eq:gsum-EM} and grouping the basis states of the ket by the momentum $p_\beta$ gives Eqs.~\eqref{eq:assembled} and \eqref{eq:assembled-EM} of the main text.
The sums over $K_n^3$ and $K_m^3$ take the form of Eq.~\eqref{eq:Tstripped}.
For a given \(L\), \(\lambda\) is further restricted by the rank and parity conditions.
The rank condition comes from the sum over $K_n^3$ in Eqs.~\eqref{eq:gsum} and \eqref{eq:gsum-EM}.
This sum vanishes unless the product of $C^\lambda$, $C^L$, and $(\sigma)^{\mathcal D}$ contains a spherical tensor of rank $0$ for $H_E$ and of rank $1$ for $E_M$.
By the triangle rule, the rank condition requires $|\lambda-L|\le\mathcal D$ for $H_E$ and $|\lambda-L|\le\mathcal D+1$ for $E_M$.
The parity condition comes from the $3j$ symbol in Eq.~\eqref{eq:elementary}.
This $3j$ symbol vanishes unless $l'+\kappa+l$ is even.
Of the operators in Eqs.~\eqref{eq:factorized} and \eqref{eq:factorized-EM}, only $C^\lambda$ and $C^L$ change $l$, so $l'-l_c$ and $\lambda+L$ are both even or both odd.
The bra and the ket basis states have the same parity.
For $\mathcal D=0$ the components $c'$ and $c$ are both upper or both lower, so $l'-l_c$ is even.
For $\mathcal D=1$ one of the two components is upper and the other is lower, so $l'-l_c$ is odd.
The parity condition therefore requires $\lambda+L$ to be even for $\mathcal D=0$ and odd for $\mathcal D=1$.

For both distributions, the rank condition and the parity condition together allow only Eq.~\eqref{eq:selection}.

\section{Truncation of the multipole sum}\label{app:lmax}

This appendix shows that the truncation $L\le L_{\max}=16$ of Sec.~\ref{sec:kernels} changes the computed distributions by much less than the accuracy of the model.

The multipole $L$ enters the distributions through the spherical Bessel function $j_L(|\bm\Delta|r)$ of Eq.~\eqref{eq:rayleigh}.
As a function of $r$, $j_L(|\bm\Delta|r)$ is close to zero until $|\bm\Delta|r$ approaches $L$.
It then rises steeply and reaches its first maximum slightly above $|\bm\Delta|r=L$.
The basis states vanish outside the box $r\le D$.
Let $r_L$ be the smallest $r$ at which $j_L(|\bm\Delta|r)$ equals half of its first maximum.
We call a truncation conservative if it keeps every $j_L$ with $r_L\le D$.
With $|\bm\Delta|=\sqrt{-t}$ and $D=30/M$, the conservative truncation at $t=-0.30$~GeV$^2$ is $L_{\max}=47$.

\begin{figure*}[!tbp]
\includegraphics[width=\textwidth]{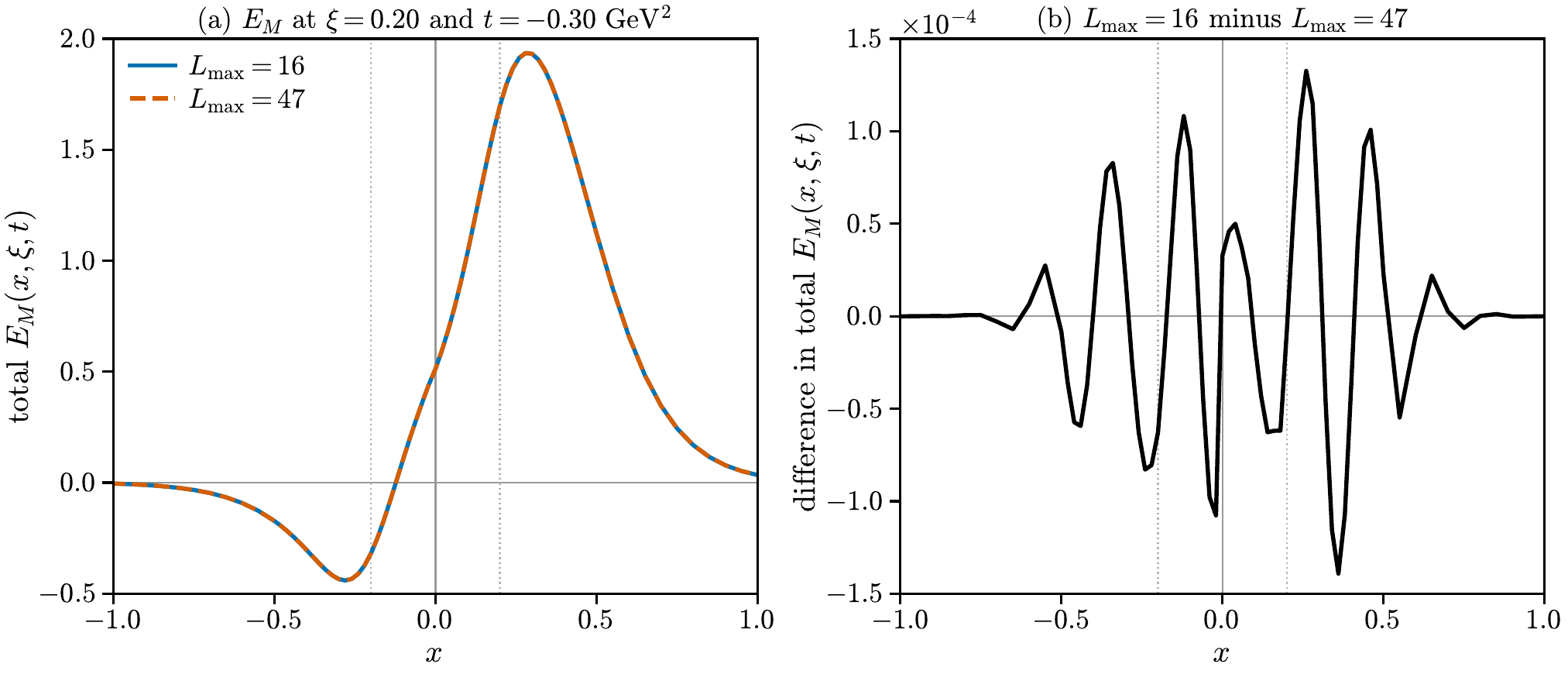}
\caption{Total $E_M(x,\xi,t)$ at $\xi=0.20$ and $t=-0.30$~GeV$^2$ with the sums over $L$ truncated at two values of $L_{\max}$.
(a) $L_{\max}=16$ (solid) and $L_{\max}=47$ (dashed).
(b) The result for $L_{\max}=16$ minus the result for $L_{\max}=47$.
The vertical dotted lines mark $x=\pm\xi$.}
\label{fig:lmax}
\end{figure*}

Figure~\ref{fig:lmax} compares $E_M$ at $\xi=0.20$ and $t=-0.30$~GeV$^2$ computed with $L_{\max}=16$ and with $L_{\max}=47$.
All other settings are those of Secs.~\ref{sec:settings} and \ref{sec:kernels}.
The two results differ by at most $1.4\times10^{-4}$, which is $0.007\%$ of the maximum of $E_M$.
The moments $\int dx\,x^nE_M$ with $n=0$, $1$, and $2$ differ by less than $4\times10^{-7}$.
The CQSM rests on the large-$N_c$ limit, and its results carry corrections of relative order $1/N_c$.
The truncation error is far below this level.
The truncation error is small because the profile and the valence state are localized at small $r$.
At $t=-0.30$~GeV$^2$, the multipoles with $L>16$ contribute mainly at $r>r_{16}=5.6$~fm, where $|F(r)|<5\times10^{-3}$.
There the sea contributions computed with $F$ and with $F=0$ in Eq.~\eqref{eq:Qreg} nearly cancel.
We therefore use $L_{\max}=16$.

A calculation with the conservative truncation is also feasible.
The cost of a kinematic point grows linearly with $L_{\max}$ (Sec.~\ref{sec:kernels}).

\bibliography{paper_gpd_xit}

\end{document}